\RequirePackage{amsmath,amsbsy,amsfonts,amssymb,enumerate,mathtools}
\RequirePackage[2018-12-01]{latexrelease}
\documentclass{gGAF2e}

\usepackage{epsfig,graphicx}
\usepackage{supertabular}
\usepackage{arydshln}

\usepackage{xcolor}
\newcommand{\gr}[1]{{\color{gray}#1}}

\begin{document}

\jvol{00} \jnum{00} \jyear{2012} 

\markboth{J.A.~Abbate \& J.M.~Aurnou}{Geophysical \& Astrophysical Fluid Dynamics}

\title{Rotating convective turbulence in moderate to high Prandtl number fluids}

\author{Jewel A.~Abbate$^{\ast}$\thanks{$^\ast$Corresponding author. Email: jewelabbate@ucla.edu
\vspace{6pt}} and Jonathan M. Aurnou\\\vspace{6pt}  Department of Earth, Planetary, and Space Sciences, University of California, Los Angeles, CA, USA\\\vspace{6pt}\received{November 3, 2023} }

\maketitle

\begin{abstract}
Rotating convective turbulence is ubiquitously found across geophysical settings, such as surface and subsurface oceans, planetary atmospheres, molten metal planetary cores, magma chambers, magma oceans, and basal magma oceans. Depending on the thermal and material properties of the system, buoyant convection can be driven thermally or compositionally, where a Prandtl number ($Pr = \nu / \kappa_i$) defines the characteristic diffusion properties of the system, with $\kappa_i = \kappa_T$ representing thermal diffusion and $\kappa_i = \kappa_C$ representing chemical diffusion. These numbers vary widely for geophysical systems; for example, the liquid iron undergoing thermal-compositional convection in Earth’s core is defined by $Pr_{T} \approx 0.1$ and $Pr_{C} \approx 100$, while a thermally-driven liquid silicate magma ocean is defined by $Pr_{T} \approx 100$. Currently, most numerical and laboratory data for rotating convective turbulent flows exists at $Pr = O(1)$; high $Pr$ rotating convection relevant to compositionally-driven core flow and other systems is less commonly studied. Here, we address this deficit by carrying out a broad suite of rotating convection experiments made over a range of $Pr$ values, employing water and three different silicone oils as our working fluids ($Pr = $ 6, 41, 206, and 993). Using measurements of flow velocities (Reynolds, $Re$) and heat transfer efficiency (Nusselt, $Nu$), a baroclinic torque balance is found to describe the turbulence regardless of Prandtl number so long as $Re$ is sufficiently large ($Re \gtrsim 10$). Estimated turbulent scales are found to remain close to onset scales in all experiments, a result that may extrapolate to planetary settings. Lastly, we use our data to build $Pr$-dependent predictive nondimensional and dimensional scaling relations for rotating convective velocities that can be applied across a broad range of geophysical fluid dynamical settings.


\end{abstract}

\section{Introduction}

Flows in geophysical and astrophysical fluid systems are often subject to convective and rotational forces that generate turbulent fluid dynamics. In most cases, the conditions of these environments are complex and extreme, with little to no direct observability. Consequently, they are studied using simplified models of rotating convection, with the aim of characterising flow behaviours fundamental to these geophysical systems that can ultimately be extrapolated to planetary scales.

The canonical model of rotating Rayleigh-Bénard convection (RBC) is often employed to investigate these types of flows. In rotating RBC, a fluid layer is heated from below while cooled from above and rotated about a vertical axis. Three dimensionless control parameters describe the system: the Rayleigh number, the Ekman number, and the Prandtl number. The Rayleigh number, $Ra$, describes the strength of thermal buoyancy effects relative to viscous and thermal diffusion and is defined as 
\begin{equation}
    Ra = \frac{\alpha g \Delta T H^3 } {\nu \kappa},
\end{equation}
where $\alpha$ is the thermal expansivity, $g$ is gravitational acceleration, $\Delta T$ is the bottom-to-top vertical temperature difference, $H$ is the fluid layer height, $\nu$ is kinematic viscosity, and $\kappa$ is thermal diffusivity. The Ekman number, $E$, describes the non-dimensional rotation period and is defined as 
\begin{equation}
    E = \frac{\nu}  {2\varOmega H^2},
\end{equation}
where $\varOmega$ is the angular rotation rate. Systems with $E \ll1$ experience significant rotational forcing relative to viscous diffusion, while $E = \infty$ defines a non-rotating system. The Prandtl number, $Pr$, describes the diffusive properties of the working fluid and is defined as
\begin{equation}
    Pr = \frac{\nu} {\kappa}.
\end{equation}
The value of $Pr$ varies greatly across natural low $E$ systems. For example, plasma within the solar convection zone is defined by an ultra low $Pr \approx 10^{-6}$ \cite[]{schumacher2020colloquium, garaud2021journey}, the liquid iron in Earth’s core by low $Pr \approx 10^{-1}$ \cite[]{roberts2013genesis}, salty ocean water is estimated to have $Pr \approx 10^{1}$ \cite[]{soderlund2019ocean}, and silicate magmas can range from moderate $Pr \approx 10^{1}$ to large $Pr \approx 10^{5}$ (or even higher depending on crystal fraction) \cite[]{lesher2015thermodynamic}.

In numerical and laboratory surveys of rotating RBC, wide ranges of $Ra$ and $E$ spanning multiple orders of magnitude are often considered. However, $Pr$ is typically fixed at $O(1)$. Numerically, this is due to the reduced cost of resolving  dynamics spatially and temporally compared to smaller or larger $Pr$ values \cite[]{horn2017prograde}. Experimentally, water ($Pr \simeq 6$) is a low cost, easily handled fluid for convection studies and is relatively close to the $Pr$ value of most simulations, providing opportunity for comparative analysis. Despite this convenience, many geophysical systems reside at more extreme values of $Pr$. Some low $Pr$ rotating convection studies have been performed using liquid metals ($Pr \simeq 10^{-2}$) \cite[]{king2013turbulent, horn2017prograde, aurnou2018rotating, guervilly2019turbulent, vogt2021oscillatory, grannan2022experimental}, but high $Pr$ studies have primarily focused on non-rotating RBC systems with viscous fluids like oils and glycerol ($Pr \simeq 10^{3}$) \cite[e.g.,][]{horn2013non, li2021effects}. High $Pr$ has generally been omitted from rotating surveys under the assumption that the large viscosity inhibits rotational effects. However, there are several high $Pr$ geophysical systems that operate at rapid enough rotation rates (i.e. sufficiently low $E$) that rotational effects are dominant. We briefly present a number of high $Pr$ geophysical systems below. 

The rotating convective turbulence within Earth’s liquid iron outer core is responsible for sustaining the planet’s global scale magnetic field. This system is convecting both thermally and compositionally as the crystallising inner core releases latent heat and light elements into the outer core \cite[]{jones2011planetary}. Using the properties of iron at core conditions, the thermal Prandtl number and chemical Prandtl number (also called the Schmidt number) are estimated to be $Pr_{T} = \nu/\kappa_T \approx 0.1$ \cite[]{roberts2013genesis} and $Pr_{C} = \nu/\kappa_C \approx 100$ \cite[]{posner2017experimental}, respectively. Many laboratory and numerical models discount the difference in thermally versus compositionally driven turbulence, arguing that turbulent mixing effectively normalises the diffusion properties of the core \cite[]{roberts2012theory}, and thus use a single effective Prandtl number of $Pr \approx 1$. However, recent work has suggested that compositional variations may be the dominant driver of buoyant convection in the core \cite[]{o2016powering, driscoll2019geodynamo, zhang2020reconciliation}, which would raise the effective Prandtl number to a value closer to $Pr_C$. The Ekman number for the outer core is small at $E = 10^{-15}$, defining Earth’s core as a predominantly high $Pr$, low $E$ system.

Early in its evolutionary history, Earth experienced extensive melting due to large impacts which created deep, global-scale magma oceans \cite[]{melosh1990giant}. During this time, the liquid silicate magma had a low crystal fraction and therefore low viscosity, while Earth rotated rapidly \cite[]{lock2017structure}, such that $Pr_T \approx 1-100$ and $E \approx 10^{-14}$ \cite[]{solomatov2000fluid}. At these parameters, rotation likely played an important role in ocean dynamics prior to mantle solidification \cite[e.g.,][]{maas2019dynamics}. It has been further proposed that the preferential solidification of the magma ocean from the outside-in created a long-lived basal magma ocean that may have been capable of generating Earth’s ancient magnetic field \cite[]{ziegler2013implications, stixrude2020silicate}. This dynamo system is controlled via compositional convection as dense iron is released from the solidifying silicate magma, which then sinks toward the core. The parameters for this system would vary over time, but are approximated as $Pr_C \approx 10^4$ and $E \approx 10^{-13}$ \cite[]{solomatov2000fluid, ziegler2013implications}. A basal magma ocean has also been proposed to exist in modern-day Venus, which would be at a comparable $Pr_C$ and $E \approx 10^{-9}$ \cite[]{o2020venus}. Additionally, most magma chambers on Earth are approximated with $Pr_T \approx 10^{5}$ and $E \approx 10^{-5}$, although subject to the age and size of the chamber \cite[]{zambra2022temporal}.

Sub-glacial lakes and their application as an analog for the sub-surface oceans of icy moons are also of great interest. Lake Vostok in Antarctica, the largest sub-glacial lake on Earth, is buried under a 4 km thick sheet of ice. This ice-water system is at $Pr_T \approx 13$ and $E \approx 10^{-8}$ \cite[]{studinger2004estimating}, parameters at which rotational effects are likely present. Rotation could affect circulation patterns that transport mass and energy through the lake, which in turn affect the potential for life in an otherwise extreme environment \cite[]{couston2021turbulent}. Similar effects likely exist in the sub-surface oceans of icy moons. Europa’s ocean, for example, is characterised by $Pr_T \approx 11$ and $E \approx 10^{-12}$ \cite[]{soderlund2019ocean}, not dissimilar from Lake Vostok. However, these extraterrestrial oceans are likely to be salty, such that both chemical and thermal fluxes control their rotating thermo-compositional convective flows \cite[]{bire2022exploring, kang2022does}.

\begin{figure}
\begin{center}
\resizebox{1.0\linewidth}{!}{\includegraphics{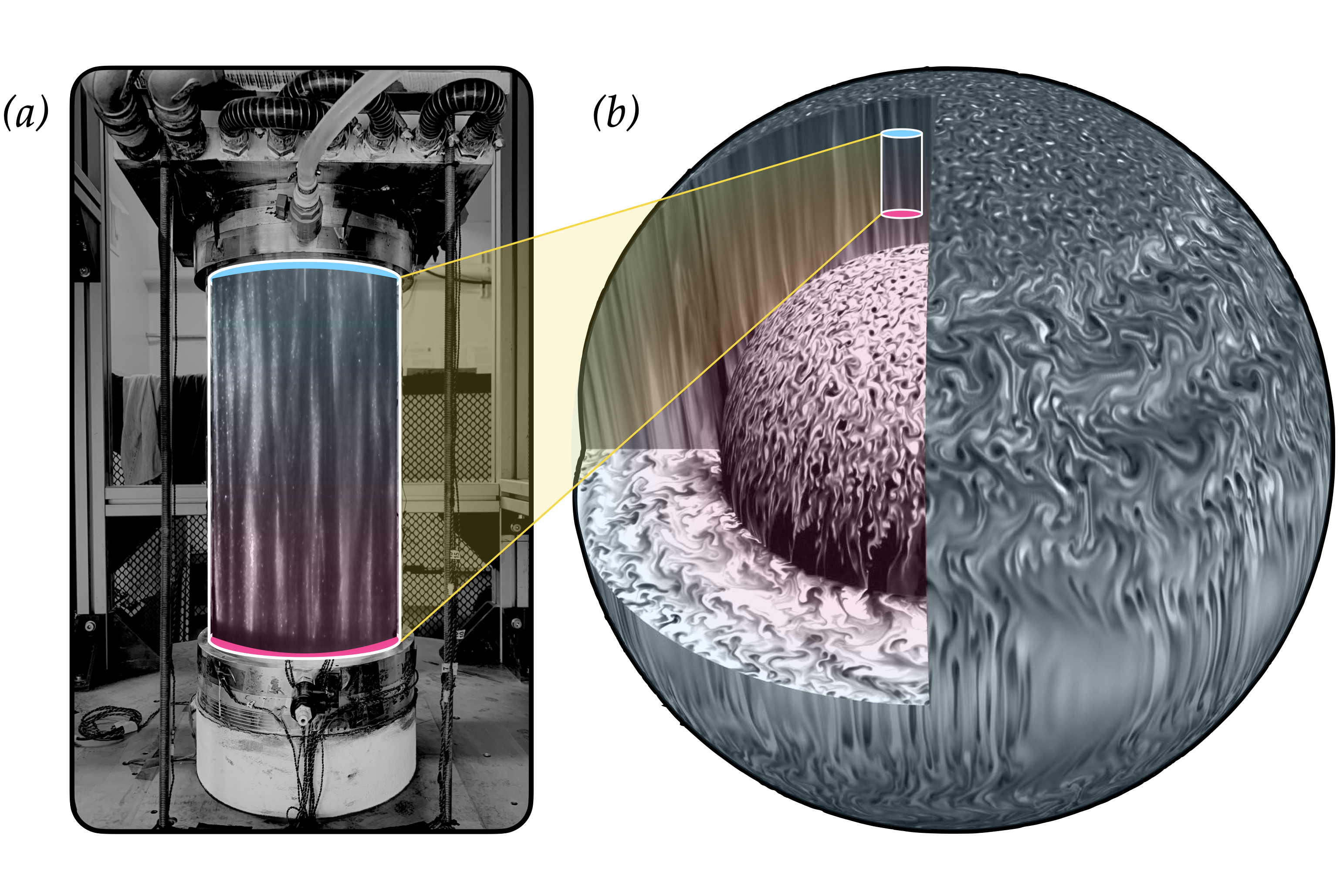}}
\caption{Schematized laboratory and numerical visualisation. Inset on the left panel is a flow visualisation done in water seeded with Kalliroscope particles ($Ra = 2 \times 10^9$, $E = 2 \times 10^{-6}$, $Pr = 6$, $Ro_c = 0.04$). Note that the image is from a $H$ = 20 cm tank and has been stretched to fit the displayed $H$ = 40 cm tank for the purpose of demonstration. The right panel displays a numerical simulation adapted from \cite{gastine2016scaling} ($Ra = 3 \times 10^9$, $E = 10^{-6}$, $Pr = 1$, $Ro_c = 0.05$). (Colour  online)}%
\end{center}
\end{figure}

The goal of this work is to investigate the effect of rotation on moderate to high Prandtl number convection, and to generate predictive scaling arguments that can be used across a broad range of natural flow systems. To do this, we perform laboratory experiments that utilise a rotating RBC set-up, in which a cylindrical cell filled with either water (moderate $Pr$) or silicone oil (high $Pr$) is heated from below, cooled from above, and rotated around its vertical axis. This set-up, shown in figure 1a, is roughly representative of a local volume in the high latitude polar regions of the liquid layers of planetary bodies \cite[e.g.,][]{gastine2023latitudinal}. The schematic in figure 1b shows a numerical simulation of Earth’s core from \cite{gastine2016scaling} compared to a laboratory visualisation of rotating convective flow in water from this study.

We study thermally-driven rotating convection in fluids with thermal $Pr$ greater than unity. The experiments can therefore be applied to $Pr_T \gtrsim 1$ geophysical systems, but may also be treated as proxies for $Pr_C \gtrsim 1$ compositionally-driven rotating convection \cite[cf.][]{classen1999blob, calkins2012effects, bouffard2019chemical}. Note that doubly-diffusive rotating convection is not considered here \cite[e.g.,][]{moll2017double, guervilly2022fingering, fuentes2023heat}, as we are assuming there is only one dominant buoyant driver of the convective flow.

Systematic measurements of vertical convective velocities, internal temperature fluctuations, and heat transfer efficiency are made across four orders of magnitude in $Ra$ ($10^8 \lesssim Ra \lesssim 10^{12}$), three orders in $E$ ($10^{-4} \gtrsim E \gtrsim 10^{-7}$), and three orders in $Pr$ ($6 \lesssim Pr \lesssim 10^{3}$). Ultimately we find, across all Prandtl numbers tested, that the vertical convective velocities are controlled by a local torque balance between the Coriolis and buoyancy terms in the vorticity equation, similar to \cite{aubert2001systematic, king2013turbulent, schwaiger2019force, long2020scaling}. We additionally find that the heat transfer is controlled by the boundary layers in all cases \cite[cf.][]{kolhey2022influence, oliver2023small, hawkins2023laboratory}, which contrasts with $Pr \ll 1$ bulk-dominated systems that are often considered in studies of planetary core dynamics \cite[e.g.,][]{king2013turbulent, vogt2021oscillatory}. 

\section{Scaling predictions for turbulent convective flows}

\subsection{Measured non-dimensional parameters}
Three measured parameters are used to characterise the flows in our rotating RBC experiments. These are the Nusselt number, $Nu$, which defines the total heat flux relative to conduction, Reynolds number, $Re$, which defines inertial advection relative to viscous diffusion, and $\theta / \Delta T$, which is the local dimensional temperature perturbation ($\theta$) normalised by the vertical temperature difference ($\Delta T$). Together, these are given as 
\begin{subequations}
\begin{gather}
    Nu = \frac{qH }{ k \Delta T}, \qquad 
    Re = \frac{ u H }{\nu}, \qquad 
    \frac{\theta }{ \Delta T},
    \tag{\theequation a-c}
\end{gather}
\end{subequations}
where $u$ is the dimensional characteristic flow velocity and $q$ is the total vertical heat flux through the system. These three quantities are related through the definition for the Nusselt number:
\begin{equation}
    Nu = \frac{q_{total}}{q_{cond}} = \frac{q_{cond} +q_{conv}}{q_{cond}} = 1 + \frac{q_{conv}}{q_{cond}},
\end{equation}
where $q_{cond}=k \Delta T / H$ and $q_{conv}=\rho c_P \langle u_z \theta \rangle $ are the conductive and convective heat flux, respectively. Here, $\rho$ is the fluid density, $c_{P}$ is the specific heat capacity, $u_z$ is the vertical component of velocity, and $\langle ... \rangle$ denotes a turbulent time and volume average \cite[]{siggia1994high}. In this study, we acquire point-wise data in time and therefore approximate $ \langle u_z \theta \rangle \sim u_{z,rms} \theta_{std}$, where `rms' is a root-mean-square time-average and `std' indicates a standard deviation in time (see sections 3.2 and 3.4). The rms value is used to represent the velocity fluctuation since the mean rotating convection velocity values are close to zero. Re-writing (5) for $Nu-1$ then gives
\begin{equation}
    Nu-1 = \frac{q_{conv}}{q_{cond}} = \frac{\rho c_P \langle u_z \theta \rangle}{k \Delta T/H} \sim RePr \frac{\theta}{\Delta T},
\end{equation}
which can be further re-arranged to yield
\begin{equation}
    \frac{\theta}{\Delta T} \sim \frac{Nu-1}{RePr}.
\end{equation}
The relation in (6) is tested against our measurements in section 4.2, while the relation in (7) is implemented here to predict $\theta / \Delta T$ scaling behaviour.

We additionally calculate the Rossby number, $Ro$, for rotating cases. This parameter defines the strength of inertial advection relative to Coriolis accelerations and is given as
\begin{equation}
    Ro = \frac{u}{2 \varOmega H}.
\end{equation}
This parameter is often compared to the convective Rossby number, $Ro_c$, which defines the strength of thermal buoyancy relative to Coriolis \citep{aurnou2020connections} and is given as 
\begin{equation}
    Ro_c = \sqrt{\frac{Ra E^2}{Pr}}.
\end{equation}

The goal of the following sections, 2.2 through 2.4, is to use the momentum, vorticity, and thermal energy equations to generate scaling expectations for the measured parameters ($Nu$, $Re$, $\theta/\Delta T$), which can then be compared to our experimentally-obtained measurements.

\subsection{Non-rotating convection}
Non-rotating RBC scaling estimates for $Re$ and $Nu$ can be derived from the expected boundary layer and bulk flow dynamics described by the momentum and thermal energy equations. 

Moderate $Pr$ fluids, such as water, are subject to turbulent mixing that generates a predominantly isothermal fluid bulk, forcing all temperature changes to occur within the thermal boundary layers. The heat transfer is then predicted to follow the classical heat transfer scaling, $Nu \sim Ra^{1/3}$ \cite[]{malkus1954heat, ahlers2009heat, king2012heat, cheng2015laboratory}. For turbulent momentum transfer, a balance between inertia and thermal buoyancy is expected to yield $Re \sim \left(Ra/Pr\right)^{1/2}$. The corresponding $\theta / \Delta T$ is then estimated using (7), which completes the following predictions:
\begin{subequations}
\begin{gather}
    Nu \sim Ra^{1/3}, \qquad 
    Re \sim \left(Ra/Pr\right)^{1/2}, \qquad 
    \frac{\theta}{ \Delta T} \sim Ra^{-1/6} Pr^{-1/2}.
    \tag{\theequation a-c}
\end{gather}
\end{subequations}

Large $Pr$ fluids, such as high-viscosity oils, are subject to strong viscous effects and thickened mechanical boundary layers $\delta_{\nu}$, following $\delta_{\nu}/H = 0.25/\sqrt{Re}$ \cite[e.g.,][]{schlichting2016boundary}. \cite{shishkina2017scaling} use dissipation arguments to show that the increased contribution from the boundary layer yields a boundary-layer-dominated large-Pr RBC regime with an expected Reynolds number scaling of $Re \sim Ra^{2/3}/Pr$. \cite{wen2020steady} show that the same scaling prediction is found in the asymptotic limit of steady 2D RBC at both small and large Prandtl numbers ($10^{-2} \leq Pr \leq 10^2$). Thermally, the interior is expected to remain isothermal and therefore follow (10a). Combining these with (7) provides the following predictions for large $Pr$ flows:
\begin{subequations}
\begin{gather}
    Nu \sim Ra^{1/3}, \qquad 
    Re \sim \frac{Ra^{2/3}}{Pr}, \qquad 
    \frac{\theta}{\Delta T} \sim Ra^{-1/3}.
    \tag{\theequation a-c}
\end{gather}
\end{subequations}

\subsection{Rotating heat transfer}
A prominent effect of rotating the RBC system about its vertical axis is the suppression of axial motions. This then greatly alters the critical Rayleigh number, $Ra_c$, defined as the value of $Ra$ (i.e. the thermal forcing) at which convection will onset. For $Pr>0.68$, plane layer rotating convection is predicted to onset via steady ($S$) motions at  
\begin{equation}
    Ra_{c,S} = 8.7 E^{-4/3}
\end{equation}
\cite[]{chandrasekhar1961, kunnen2021geostrophic}, where the Ekman number, $E$, describes the system's nondimensional rotation period. However, if $Pr\leq0.68$, rotating convection onsets as thermal-inertial oscillations ($O$), which are predicted to be present in a horizontally infinite plane at
\begin{equation}
    Ra_{c,O} = 17.4 \left( E/Pr \right)^{-4/3}
\end{equation}
\cite[]{chandrasekhar1961, julien1998new}. This low $Pr$ prediction is not relevant to the data shown here, but arises in section 5.1. The convective supercriticality, $\widetilde{Ra}$, then defines how far past the onset of convection a given system is. It is defined here as
\begin{equation}
    \widetilde{Ra} = Ra/Ra_c,
\end{equation}
where $Ra_c=Ra_{c,S}$ if $Pr>0.68$ and $Ra_c=Ra_{c,O}$ if $Pr\leq0.68$.

\subsection{Local torque balance in rotating convection}
Local turbulent motions in the fluid bulk are assumed to be controlled by a baroclinic torque balance between the Coriolis and thermal buoyancy terms of the vorticity equation \cite[cf.][]{aubert2001systematic, king2013turbulent, jones2015thermal}, which is given as
\begin{equation}
    2 \mathbf{\varOmega} \cdot{} \nabla \mathbf{u} = \nabla \times (\alpha g \theta \mathbf{\hat{z}}).
\end{equation}
Scaling $\nabla\sim1/H$ in the Coriolis term and $\nabla\sim1/\ell$ in the buoyancy term yields
\begin{equation}
    \frac{2 \varOmega u} { H} \sim \frac{g \alpha \theta }{\ell},
\end{equation}
where $\ell$ is the horizontal scale of convection. Solving for $u$ returns the characteristic convective velocity,
\begin{equation}
    u \sim \frac {g \alpha \theta }{2 \varOmega} \frac{H}{\ell}.
\end{equation}
Re-writing (17) for $\theta$, substituting it into (6), and simplifying yields the rotating Reynolds number prediction,
\begin{equation}
    Re \sim \left[ \frac{Ra (Nu-1)}{Pr^2} E \left( \frac{H}{\ell} \right) \right]^{1/2}.
\end{equation}
Alternatively, a scaling relation for $\theta/\Delta T$ can be obtained by substituting (18) for $Re$ in (7), which yields
\begin{equation}
    \frac{\theta}{\Delta T} \sim \left[ \frac{Nu-1}{RaE} \ \left( \frac{\ell}{H} \right) \right]^{1/2}.
\end{equation}
Together, these relations represent the non-dimensional velocity and temperature fluctuations predicted for rotating convective flows. They are both directly coupled to the heat transfer, $Nu$, which is measured in this study, as well as the local cross-axial length scale, $\ell$, which is not measured here. An additional balance with either viscosity or inertia is considered in order to determine estimates of $\ell$ that can be substituted into (18) and (19).

\subsubsection{Viscous-Archimedean-Coriolis (VAC) balance}
A triple balance between the viscosity, buoyancy, and Coriolis terms of the vorticity equation defines the VAC system. The corresponding length scale estimate is determined by equating the Coriolis and viscous terms,
\begin{equation}
    2 \mathbf{\varOmega} \cdot{} \nabla \mathbf{u} = \nu \nabla^2 \mathbf{\varOmega},
\end{equation}
where $\varOmega=\nabla\times\mathbf{u}$ is the vorticity. Scaling $\nabla\sim1/H$ in the Coriolis term and $\nabla\sim1/\ell$ in the viscous term yields
\begin{equation}
    \frac{2 \varOmega u} { H} \sim \frac{ \nu u }{\ell^3},
\end{equation}
which can be simplified to 
\begin{equation}
    \frac{\ell}{H} \sim E^{1/3}
\end{equation}
\cite[]{stellmach2004cartesian}. This scale also arises from linear stability theory as the critical length scale, $\ell_{crit}$, at the onset of convection. The exact relation follows $\ell_{crit}/H = 2.4 E^{1/3}$ for $Pr > 0.68$ and $\ell_{crit}/H = 2.4 (E/Pr)^{1/3}$ for $Pr \leq 0.68$ \citep{chandrasekhar1961}. The pre-factors are not carried through in (23) and (24), but are used in subsequent plots and analysis. Substituting (22) into (18) and (19) yields non-dimensional predictions of
\begin{equation}
    Re_{VAC} \sim \left[ \frac{Ra (Nu-1)}{Pr^2} \right]^{1/2} E^{1/3},
\end{equation}
\begin{equation}
    \frac{\theta_{VAC}}{\Delta T} \sim \left[ \frac{Nu-1}{RaE^{2/3}} \  \right]^{1/2}.
\end{equation}
This balance is typically argued to exist in moderate to high $Pr$ fluids under moderate $\widetilde{Ra}$ conditions \cite[]{aubert2001systematic, king2013turbulent, gastine2016scaling}.

\subsubsection{Coriolis-Inertia-Archimedean (CIA) balance}
A triple balance between the Coriolis, inertia, and buoyancy terms of the vorticity equation defines the CIA system. The corresponding length scale estimate is determined by equating the Coriolis and inertial terms,
\begin{equation}
    2 \mathbf{\varOmega} \cdot{} \nabla \mathbf{u} = \mathbf{u} \cdot{} \nabla \mathbf{\varOmega}.
\end{equation}
Scaling $\nabla\sim1/H$ in the Coriolis term and $\nabla\sim1/\ell$ in the inertial term yields
\begin{equation}
    \frac{2 \varOmega u} { H} \sim \frac{ u^2 }{\ell^2},
\end{equation}
which can be recast as
\begin{equation}
    \frac{\ell_{turb}}{H} \sim Ro^{1/2},
\end{equation}
where the turbulent length scale, $\ell_{turb}$, defines the characteristic cross-axial length scale of convection in inertially dominated systems. Substituting this estimate into (18) and (19) yields non-dimensional predictions of
\begin{equation}
    Re_{CIA} \sim \left[ \frac{Ra (Nu-1)}{Pr^2} \right]^{2/5} E^{1/5},
\end{equation}
\begin{equation}
    \frac{\theta_{CIA}}{\Delta T} \sim \left[ \frac{Nu-1}{RaE} \  \right]^{3/5} Ro_c^{2/5}.
\end{equation}
This description is predicted to hold when inertial turbulence dominates the bulk motions, but the heat transfer is still controlled by micro-scale (diffusive) boundary layer processes \cite[]{kolhey2022influence, oliver2023small, hawkins2023laboratory}.

\subsubsection{Diffusivity-free (DF) limit}
The diffusivity-free system is defined as the case in which not only momentum transfer is free of viscous effects (CIA balance), but when heat transfer is diffusion-free as well \cite[e.g.,][]{julien2012heat, aurnou2020connections, bouillaut2021experimental}. This fully inertial system approaches the inviscid limit of rapidly rotating convection, in which an unstable thermal gradient is sustained in the fluid bulk such that 
\begin{equation}
    \frac{\theta}{\Delta T} \sim \frac{\ell}{H}
\end{equation}
\cite[]{sprague2006numerical, julien2012heat, aurnou2020connections}.
Following \cite{aurnou2020connections}, this relation can be substituted for $\ell$ in (17) and simplified to return a characteristic velocity for fully diffusivity-free (DF) flows:
\begin{equation}
    u_{DF} \sim \frac {g \alpha \theta }{2 \varOmega} \frac{\Delta T}{\theta}\sim \frac {g \alpha \Delta T }{2 \varOmega}.
\end{equation}
Substituting this velocity into $Re=uH/\nu$ and utilising the relation $Ro=ReE$ gives non-dimensional scaling predictions of
\begin{subequations}
\begin{gather}
    Re_{DF} \sim  \frac{Ro_c^2}{E}, \qquad 
    Ro_{DF} \sim Ro_c^2.
    \tag{\theequation a-b}
\end{gather}
\end{subequations}
This system is still in CIA balance, so $Ro_{DF}$ can be substituted for $Ro$ in (27) to yield a specific length scale estimate:
\begin{equation}
    \frac{\ell_{DF}}{H} \sim Ro_{DF}^{1/2} \sim Ro_c.
\end{equation}
The relation in (30) then implies
\begin{equation}
    \frac{\theta_{DF}}{\Delta T} \sim Ro_c.
\end{equation}
This result can be used to extract the $Nu$ estimate for diffusivity-free flows. Setting $Re_{CIA}\sim Re_{DF}$ and solving for $Nu-1$ gives
\begin{equation}
    Nu-1 \sim Ro_c RaE \sim Pr^{-1/2} Ra^{3/2} E^2,
\end{equation}
in agreement with \cite{julien2012heat}.

While $\ell_{DF}$ is expected in the fully asymptotic system, we additionally consider the case in which the flow scale has approached this limit, but heat transfer has not. Maintaining the measured $Nu$ in (18) and (19), but substituting $\ell=\ell_{DF}$ yields
\begin{equation}
    Re_{CIA*} \sim \left[ \frac{Ra (Nu-1)}{Pr^2} \frac{E}{Ro_c} \right]^{1/2},
\end{equation}
\begin{equation}
    \theta_{CIA*}/\Delta T \sim \left[ \frac{Nu-1}{RaE} \ Ro_c \right]^{1/2},
\end{equation}
where $CIA*$ indicates a CIA scaling modified to include the asymptotic length scale.

All scaling predictions presented here are tested against experimentally-obtained $Re$, $Nu$, and $\theta/\Delta T$ data in section 4.4.

\section{Methods}

\subsection{Laboratory experiment}
Figure 2a) shows the laboratory experimental set-up. Rotating and non-rotating RBC experiments are conducted in an axially-aligned cylindrical cell of inner radius $R$ = 9.64 cm. The cell height is varied between $H$ = 19.05, 38.10, and 80.01 cm, yielding three tested aspect ratios: $\varGamma = 2R/H$ = 0.25, 0.5, and 1.0. The cell sidewall is made of acrylic (thermal conductivity $k=0.19$ W/m/K) and is wrapped in 3 cm of aerogel insulation ($k=0.015$ W/m/K) to mitigate lateral heat losses. The top and bottom plate boundaries are made of T6061 aluminum ($k=167$ W/m/K). 

\begin{figure}[b]
\begin{center}
\resizebox{1.0\linewidth}{!}{\includegraphics{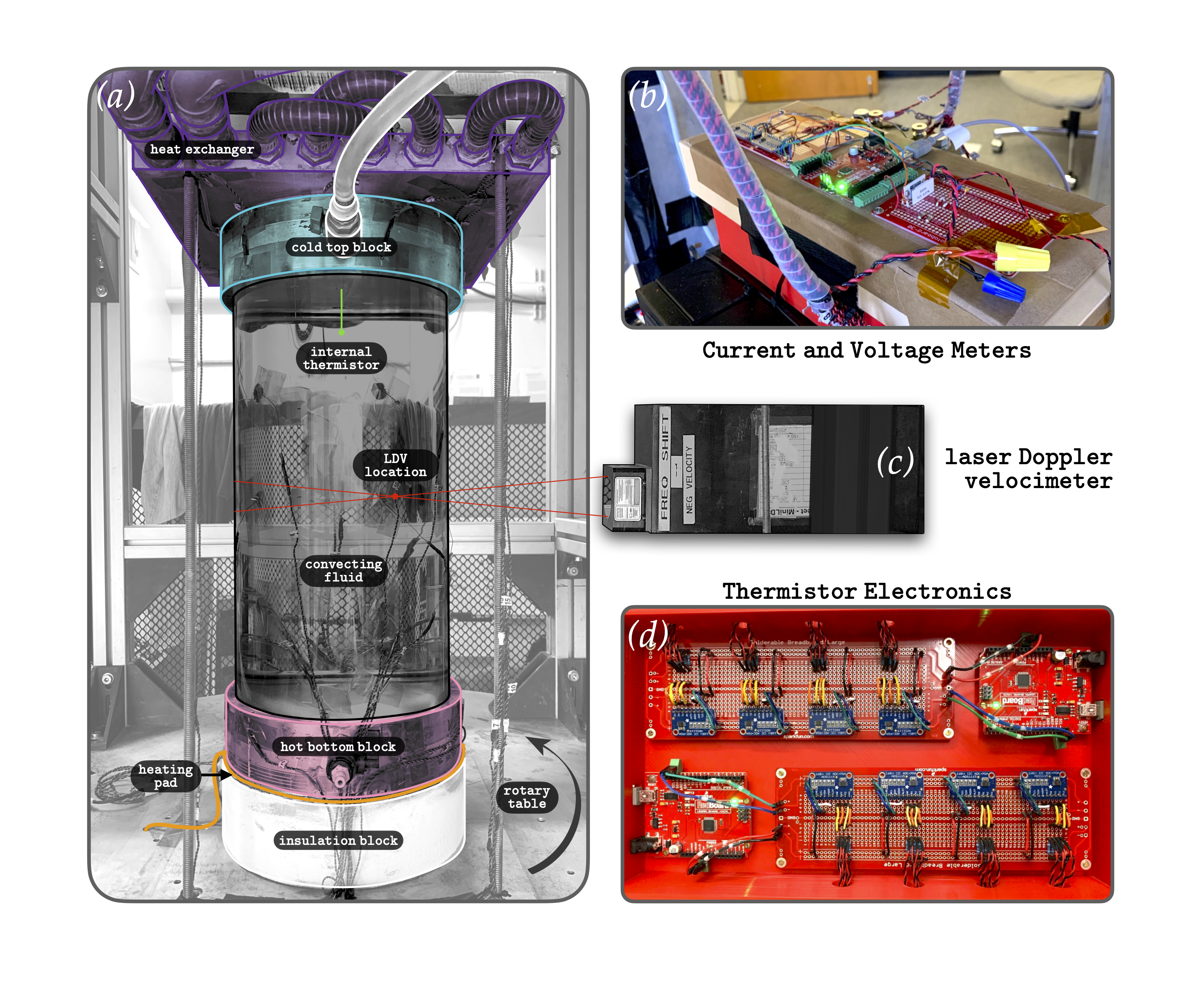}}
\caption{a) Laboratory device schematic with the H = 40 cm tank. b) Electronics for current and voltage meters. c) Laser Doppler velocimeter (LDV) positioned at the mid-plane to measure vertical velocity, $u_z$. d) Thermistor electronics measure temperature at the top and bottom boundaries, internal fluid temperature (at $d$ = 5.0 cm below the top boundary for all tanks), and temperatures along the sidewall. (Colour  online)}%
\end{center}
\end{figure}

The bottom plate sits atop a non-inductively wound heating pad, which is supplied power ranging from $P = 10$ -- 400 W. A 7.6 cm thick insulation block ($k=0.15$ W/m/K) sits below the heating pad to inhibit heat loss down into the table. The input power is set using an Arduino microcontroller programmed to maintain a constant, user-defined, bottom plate temperature. The top plate is thermo-stated by a thermally-coupled, aluminum heat exchanger plate in series with a fixed-temperature recirculating chiller. The temperature control in these experiments enables cases at fixed Rayleigh numbers across all tanks and fluids used in the study. Dimensionally, experiments are conducted with vertical temperature gradients as low as $\Delta T=2$ K and as high as $\Delta T=40$ K in water and $\Delta T=55$ K in oil. The upper end of our $\Delta T$ range is high enough to potentially introduce non-Boussinesq effects \cite[]{horn2013non, horn2014rotating, ricard2022fully}, however, the results presented in section 4 do not show any indication of a shift in behaviour at high $\Delta T$. Non-dimensionally, the experiments are conducted with Rayleigh numbers of approximately $Ra=3\times10^8$ to $Ra=2\times10^{12}$ across all experiments.

The convection cell is mounted on a rotating platform. Power is supplied from the stationary (lab) frame to the onboard electronics and heating pad through an electrical slip ring at the base of the platform. Cool water from the recirculating chiller is transported through a fluid rotary union mounted above the convection cell. The platform is programmed to rotate at rates ranging from $\varOmega=0.8$ rpm to $\varOmega=35$ rpm, enabling fixed Ekman number cases. Across all experiments, the Ekman number spans approximately $E=2\times10^{-7}$ to $E=4\times10^{-4}$. The Froude number, given by $Fr=\varOmega^2R/g$, estimates the strength of centrifugation, where the critical Froude number for large-scale convective flow is argued to be $Fr_c = \varGamma/2$ \cite[]{horn2018regimes, horn2019rotating, horn2021tornado}. The data presented herein are all at $Fr<Fr_c$, with a maximum value of $Fr/Fr_c = 0.55$. Thus, the effects of centrifugal force are not considered.
	
The working fluids are water and silicone oil of varied viscosities. The material properties (density, $\rho$ [kg/m$^3$], thermal expansivity, $\alpha$ [1/K], thermal diffusivity, $\kappa$ [m$^2$/s], and thermal conductivity, $k$ [W/m/K]) are determined using the mean temperature of the fluid, $\overline{T} = (T_{top} + T_{bot})/2$, with a temperature-dependent equation for each property. We use equations from \cite{lide2004crc} for water and equations determined via manufacturer information for the silicone oils \cite[see][]{gelest}. The oils used are the 3 cSt, 20 cSt, and 100 cSt Conventional Silicone Fluid from Gelest, Inc. At room temperature ($T=25^{\circ}$C), the Prandtl numbers of the working fluids are approximately $Pr$ = 6, 41, 206 and 993.

\subsection{Thermometry}
The thermal state of the system is determined using a series of Amphenol NTC thermistor temperature sensors placed throughout the experimental set-up. Shown in figure 2d, Arduino microcontrollers with 16-bit analog-to-digital converters (Adafruit ADS1115) are wired to 32 thermistors, each of which is custom calibrated in-house to be accurate to within $\pm$ 0.02 K.

Six of these thermistors are embedded in each top and bottom aluminum block \cite[see figures 2 in][]{king2012heat, xu2022thermoelectric}. They are spaced evenly in azimuth and are all within 2 mm of the fluid-lid interface. These temperatures are used to determine the time-averaged vertical temperature difference, given by
\begin{equation}
    \Delta T = T_{bot}-T_{top} =  \frac{1}{6}\sum_{i=1}^{6}{\overline{T}_{bot,i}}-\frac{1}{6}\sum_{i=1}^{6}{\overline{T}_{top,i}},
\end{equation}
where $\overline{T}_i$ is the time-average of the $i^{th}$ thermistor in the block. The temperature variance across each block is small at $\leq 0.1$ K for each case, such that the boundaries may be considered effectively isothermal.

An internal thermistor positioned within the fluid bulk at radial position $r$ = 0 and depth $d$ = 5.0 cm below the top boundary measures the local temperature perturbation, $\theta$. The depth of the thermistor is set by the maximum thermistor length we had available at the time of this study. The thermal and viscous boundary layers for all rotating experiments are smaller than 5 cm, however, so these measurements are assumed to sample the fluid bulk. The $\theta$ value is the standard deviation of the temperature time series:
\begin{equation}
    \theta = \sqrt{ \sum_{i=1}^{N}  \frac{(T_i-\overline{T})^2}{N} },
\end{equation}
where $T_i$ represents each measurement in the time series of length $N$ and $\overline{T}$ is the time-average of the entire steady-state portion of the time series.

Additionally, an array of thermistors is placed on the cell sidewall to monitor the presence of possible large-scale circulations \cite[]{weiss2011large} or boundary zonal flows \cite[]{zhang2021boundary}, as well as to quantify heat loss at the sidewall. One thermistor is located at the location of the heating pad for monitoring and controlling its temperature, and another thermistor is placed outside of the convection cell to assess room temperature.

\subsection{Heat flux}
The raw input power supplied to the convection cell is determined via time-averaged measurements of current, $I$, and voltage, $V$, made directly at the location of the heating pad. These are performed using an Arduino microcontroller with a current shunt and a voltage divider circuit, respectively (figure 2b). The heat flux through the base of the convection cell is then calculated as $\overline{q}_{base}=(\overline{V}\overline{I})/A_{base}$, where $A_{base}=\pi R^2$ is the active surface area in contact with the fluid. Heat lost through the cell sidewall is determined using thermistor temperature sensors located within the layers of sidewall insulation. The heat loss (in W) is estimated from the conductive temperature gradient across a cylindrical layer of insulation following
\begin{equation}
    P_{loss} = \frac{ 2 \pi k_{ins} H (T_{SW}-T_{ins}) }{log(R_{ins}/R_{SW})},
\end{equation}
where $k_{ins}$ is the thermal conductivity of the aerogel insulation;  $T_{SW}$ is the temperature measured at $R_{SW} = 10.12$ cm, the outer wall of the acrylic cylindrical tank;  $T_{ins}$ is the temperature measured after one layer of insulation (1 cm thick) at radial position $R_{ins} \simeq 11.12$ cm. Most experiments were conducted with a mean temperature comparable to the room temperature, and therefore had heat losses that were less than 1\% of the input heating power $\overline{VI}$. Nevertheless, the vertical time-averaged heat flux is calculated as
\begin{equation}
    \overline{q} = \frac{\overline{V} \overline{ I} - \overline{P_{loss}}}{\pi R^2},
\end{equation}
where the overline indicates a time-average. Since these experiments are set with a fixed temperature, the state of thermal equilibration is determined by the long period trend in heat flux. We denote a case to be equilibrated when the measured heat flux is steady for at least one hour, and continue to record data for at least two more hours. Equilibration time varies for each case, but usually occurs in one to six hours, with the lowest heat flux cases taking as long as 24 hours.

The Nusselt number, $Nu$, is calculated using the heat flux given by (41). The heat lost from the system is considered to be the dominant error in the heat flux calculation and is therefore included in the error propagation for $Nu$. The error is then calculated as:
\begin{equation}
    \delta_{Nu} = Nu \sqrt{ \left( \frac{P_{loss}}{P} \right)^2 + 2 \left( \frac{\delta_R}{R} \right)^2 + \left( \frac{\delta_H}{H} \right)^2 + \left( \frac{\delta_k}{k} \right)^2 + \left( \frac{\delta_T}{\Delta T} \right)^2},
\end{equation}
where $\delta_R = \delta_H$ = 1 mm, $\delta_k$ is considered negligible, and $\delta_T$ = 0.04 K. Values of $\delta_{Nu} / Nu$ are generally less than 1\%, but increase for low heat flux and low $\Delta T$ cases.

\subsection{Velocimetry}
An MSE first-generation UltraLDV laser Doppler velocimeter (LDV) is fixed in the rotating frame to measure the single-point vertical flow velocity, $u_z$ (figure 2c). In this velocimetry device, a split laser beam converges at the point of measurement, creating a fringe pattern that seed particles (20 $\mu m$ hollow glass spheres) traverse as they move through the fluid bulk. This causes fluctuations in the scattered light intensity, the frequency of which is recorded by the velocimeter and converted into an absolute velocity \cite[]{noir2010experimental, hawkins2023laboratory}. The root-mean-square (rms) velocity is calculated from the LDV time series via
\begin{equation}
    u_{z} = \sqrt{ \sum_{i=1}^{N} \frac{u_i^2} {N}}
\end{equation}
where $u_i$ represents each velocity measurement in the time series.

The seed particles are about 10\% denser than the working fluids ($\rho_{s}$ = 1.1 g/cm$^3$), but remain suspended due to convection. The settling velocity of the particles is estimated by $ v = (1/18) [(\rho_s-\rho_f) gd^2/(\nu \rho_f)]$ \cite[]{kriaa2022effects}, where $\rho_s$ is the density of the seed particles, $\rho_f$ is the fluid density, $d$ is the particle diameter, and $\nu$ is the fluid kinematic viscosity. The settling velocities range from $v = 0.3 \ \mu$m/s to $v = 0.02$ mm/s at room temperature, while the measured convective velocities range from 2 to 20 mm/s. The ratio of settling velocity to convective flow velocity, also called the Rouse number, $ \mathcal{R} $, is always small at $ \mathcal{R} < 0.01 $ indicating low particle inertia \cite[]{kriaa2022effects}.

Figure 3a shows an azimuthal velocity ($u_{\phi}$) time series from an isothermal calibration experiment measuring linear spin-up in water \cite[cf.][]{greenspan1963time, warn1978numerical, aurnou2018rotating}. The theoretical exponential decay profile is given by 
\begin{equation}
u_{sp} = \Delta \varOmega s \exp{(-(t-t_0)/\tau)}, 
\end{equation}
where $\Delta \varOmega = \varOmega_f - \varOmega_i$ is the impulsive increase in angular rotation velocity from $\varOmega_i$ to $\varOmega_f$ implemented at time $t_o$, and $\tau=H/(2\sqrt{\nu \varOmega_i})$ is the theoretical spin-up time \cite[]{aurnou2018rotating}. The linear spin-up profile given by (44) is overlain in blue to demonstrate the agreement between the LDV measurements and the theoretical expectation. However, the velocity signal is accompanied by a consistent level of ambient noise. Figure 3b shows the measurement with the theoretical spin-up profile subtracted out, yielding an rms noise level estimate of $(u_{\phi}-u_{sp})_{rms}$ = 1.69 mm/s. This noise sets the lowest possible rms velocity that our system can measure, and ultimately corrupts measurements too close to this limit. Figure 3c shows a probability density function (PDF) of the data shown in figure 3b, which is described by a Gaussian distribution. The noise is not easily separated from the actual flow velocity because the distribution of our rotating convective velocities are also often Gaussian [cf. Hart, Kittleman, Olsen 2002; Kunnen et al. 2010].

\begin{figure}
\begin{center}
\resizebox{1.0\linewidth}{!}{\includegraphics{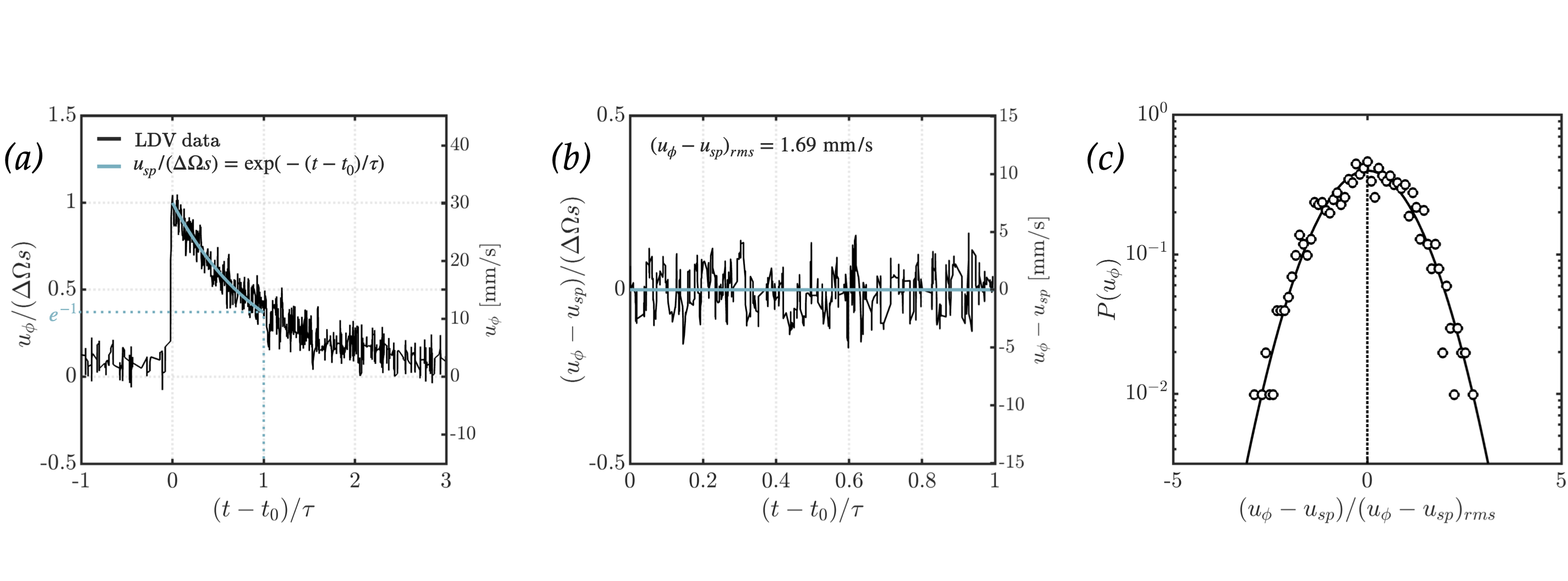}}
\caption{Isothermal linear spin-up calibration result for the LDV. Panel (a) shows the normalised $u_{\phi}$ time series with the theoretical exponential decay profile overlain in blue. The change in rotation rate is given by $\Delta \varOmega = \varOmega_f - \varOmega_i$ and the radial position of the measurement is given by $s$. The x-axis represents time after the spin-up impulse, which is normalised by the linear spin-up time given as $\tau = H/(2\sqrt{\nu \varOmega_i})$ \cite[]{aurnou2018rotating}. Panel (b) shows the normalised $u_{\phi}$ time series with the theoretical profile subtracted out for one e-folding. The rms-velocity of the remaining ambient signal is $(u_{\phi} - u_{sp})_{rms} = 1.69$ mm/s, which sets the minimum resolvable rms-velocity for our rotating system. Panel (c) shows the PDF for the ambient system noise from (b) with a reference Gaussian profile. (Colour  online)}%
\end{center}
\end{figure}

To determine an uncertainty ($\delta_{u_z}$) that accounts for the systematic noise, we utilise the signal-to-noise ratio (SNR) defined in decibel (dB) units as
\begin{equation}
    \text{SNR(dB)} = 10 \log_{10} \left[\frac{\text{signal mean square}}{\text{noise variance}} \right]=10 \log_{10} \left[ u_{z}^2 / u_{noise}^2 \right]
\end{equation}
\cite[]{shinpaugh1992signal, johnson2006signal}. The uncertainty is then calculated as $\delta_{u_{z}} / u_{z} = \text{SNR}^{-1}$, which is used to determine error on the Reynolds number:
\begin{equation}
    \delta_{Re_z} = Re_z \sqrt{ (\text{SNR}^{-1})^2 + (\delta_H/H)^2 + (\delta_{\nu}/\nu)^2},
\end{equation}
where $\delta_{\nu}$ is considered negligible. All $Re_z$ error bars shown hereafter are determined using (46). Further, we choose to omit all rotating measurements with SNR $<5.5$ dB (corresponding to $u_z<3.20$ mm/s) to retain only high quality measurements in the analysis. All measurements (including those corrupted by noise) are included in table B5 in appendix B for reference, where noisy velocities are coloured in gray. We further discuss the effect of this noise in section 5.3. The non-rotating measurements did not contain significant ambient noise, such that the measured signals were not impeded. Therefore, we include all non-rotating cases in our analysis.

\section{Results}

\subsection{Shadowgraph flow visualisation in silicone oil}
Shadowgraph imaging is used to qualitatively examine flow patterns for varied $Ra$, $E$, and $Pr$. This technique uses a point light source positioned at the sidewall to illuminate flow within the 3D cylindrical tank. The light is refracted as it travels through plumes of varying density (and therefore different indices of refraction), which creates shadows that are projected onto an imaging plane \cite[cf.][]{settles2017review}. Darker regions caused by stronger refraction indicate areas of relative higher density (lower temperature), while lighter regions indicate relative lower density (higher temperature). Silicone oil has a high coefficient of thermal expansion ($\alpha \approx 10^{-3}$ 1/K), which causes strong temperature-dependent density variations advantageous to this imaging technique. The effect is not as strong in water ($\alpha \approx 10^{-4}$ 1/K), so it is excluded here.

\begin{figure}
\begin{center}
\resizebox{1.0\linewidth}{!}{\includegraphics{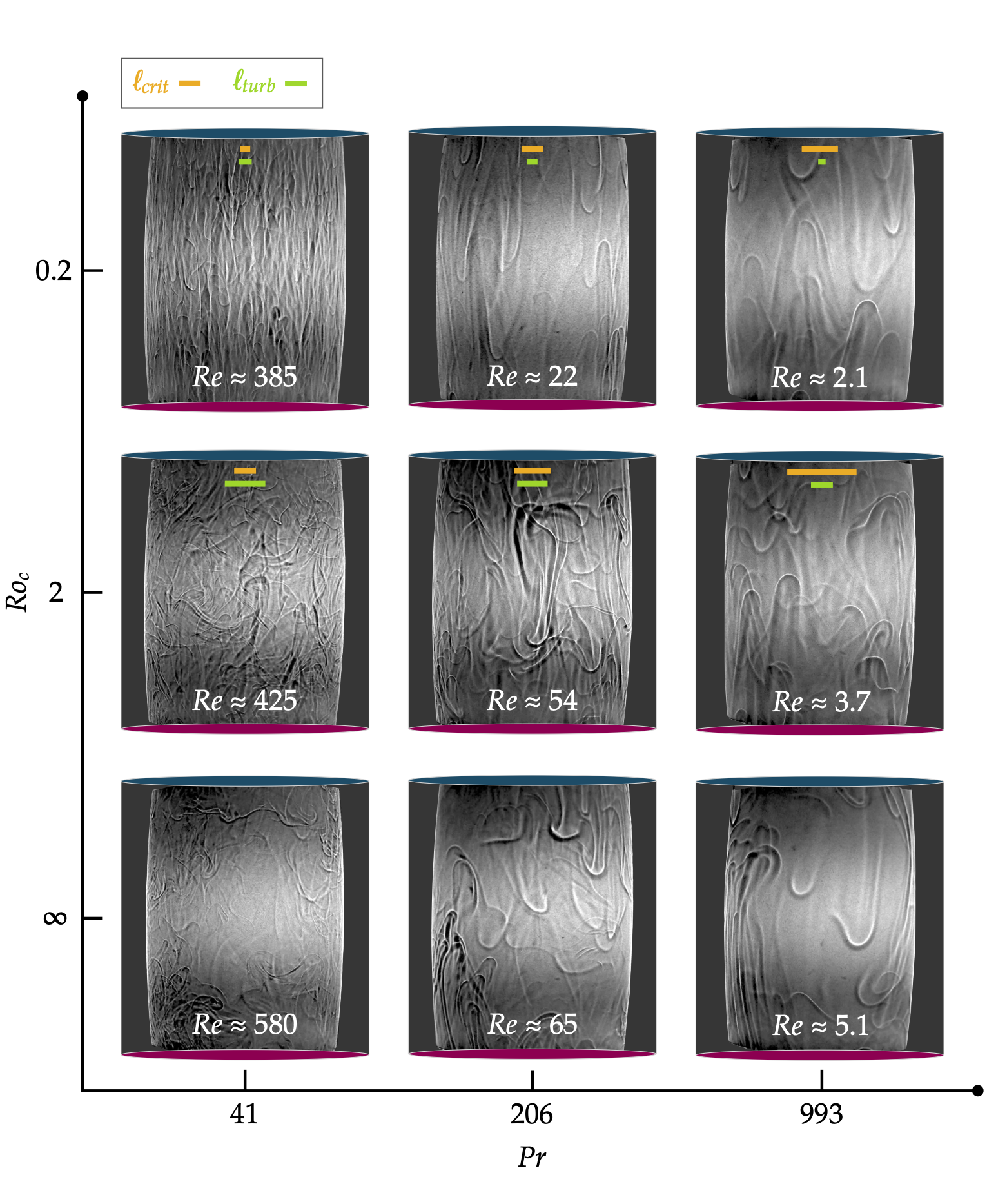}}
\caption{Laboratory visualisation diagram for the $\varGamma=1$ tank. Shadowgraph technique is used, where a point light source illuminates plumes within the 3D space of the cylindrical cell. No corrective lenses were used when imaging, therefore the displayed visuals are subject to some cylindrical distortion. Prandtl and convective Rossby numbers are approximated here; exact values are given in table B1. Reynolds numbers were not directly measured while obtaining the flow images, so their values are instead calculated using the best fit equations from table B2 for non-rotating cases and table B3 for rotating cases. Orange and green horizontal lines represent the approximate width predicted by the onset and turbulent length scales, respectively. (Colour  online)}%
\end{center}
\end{figure}

Figure 4 shows $Ro_c$ versus $Pr$ images from the $\varGamma=1$ tank. Corresponding case parameters are presented in table B1. The images qualitatively demonstrate that as $Ro_c$ decreases (and the Coriolis force begins to dominate over buoyancy), the flow becomes increasingly organised into axially-aligned structures. For $Ro_c \gtrsim 1$, the Coriolis force has a weak effect and buoyancy-driven plumes move more freely in the three-dimensional space of the tank. In the complete absence of rotation ($Ro_c=\infty$), a large-scale-circulation (LSC) defines the flow across all $Pr$, with a distinct upwelling and downwelling on opposite sides of the tank and weaker motions at the center \cite[cf.][]{shang2003measured, xi2004laminar, huang2016effects, li2021effects}. Further, as $Pr$ increases with constant $Ro_c$, the flow becomes less turbulent due to the increased kinematic viscosity (decreased $Re$), with plumes maintaining a more coherent shape for a longer period of time as they traverse the vertical length of the tank. 
	
The rotating cases in figure 4 ($Ro_c<\infty$) are annotated with horizontal bars indicating the length scale predictions of $\ell_{crit}=2.4E^{1/3}H$ and $\ell_{turb}=Ro^{1/2}H$. The Rossby number used in the $\ell_{turb}$ estimate is calculated using the experimentally-derived scaling relations presented in tables B2 and B3 because we did not simultaneously perform shadowgraph and laser Doppler velocimetry. These bars do not account for optical distortion of plumes within the cylindrical tank, but they qualitatively estimate these length scale predictions in the experiments. The most notable difference between the two predictions is that $\ell_{crit}$ increases and $\ell_{turb}$ decreases with increasing $Pr$ at constant $Ro_c$. Visually, the plumes appear to increase in width as $Pr$ is increased, consistent with the $\ell_{crit}$ trend. However, both $\ell_{crit}$ and $\ell_{turb}$ reasonably match the plume width in many cases, making them difficult to distinguish here. The ratio $\ell_{turb}$/$\ell_{crit}$ is included in table B1, the values of which are near one for all cases included in the diagram. The implications of this approximate length scale equivalence are discussed in section 5.1. 

\subsection{Simultaneous measurements of $\theta$, $Nu$, and $Re$}
Measurements of $\theta$, $Nu$, and $Re_z$ were acquired independently of each other, but can be related through the equation given by (6). Figure 5 checks the accuracy of (6) for the data acquired here by showing $Nu-1$ versus $Re_z Pr \theta/\Delta T $. 

\begin{figure}[b]
\begin{center}
\resizebox{1.0\linewidth}{!}{\includegraphics{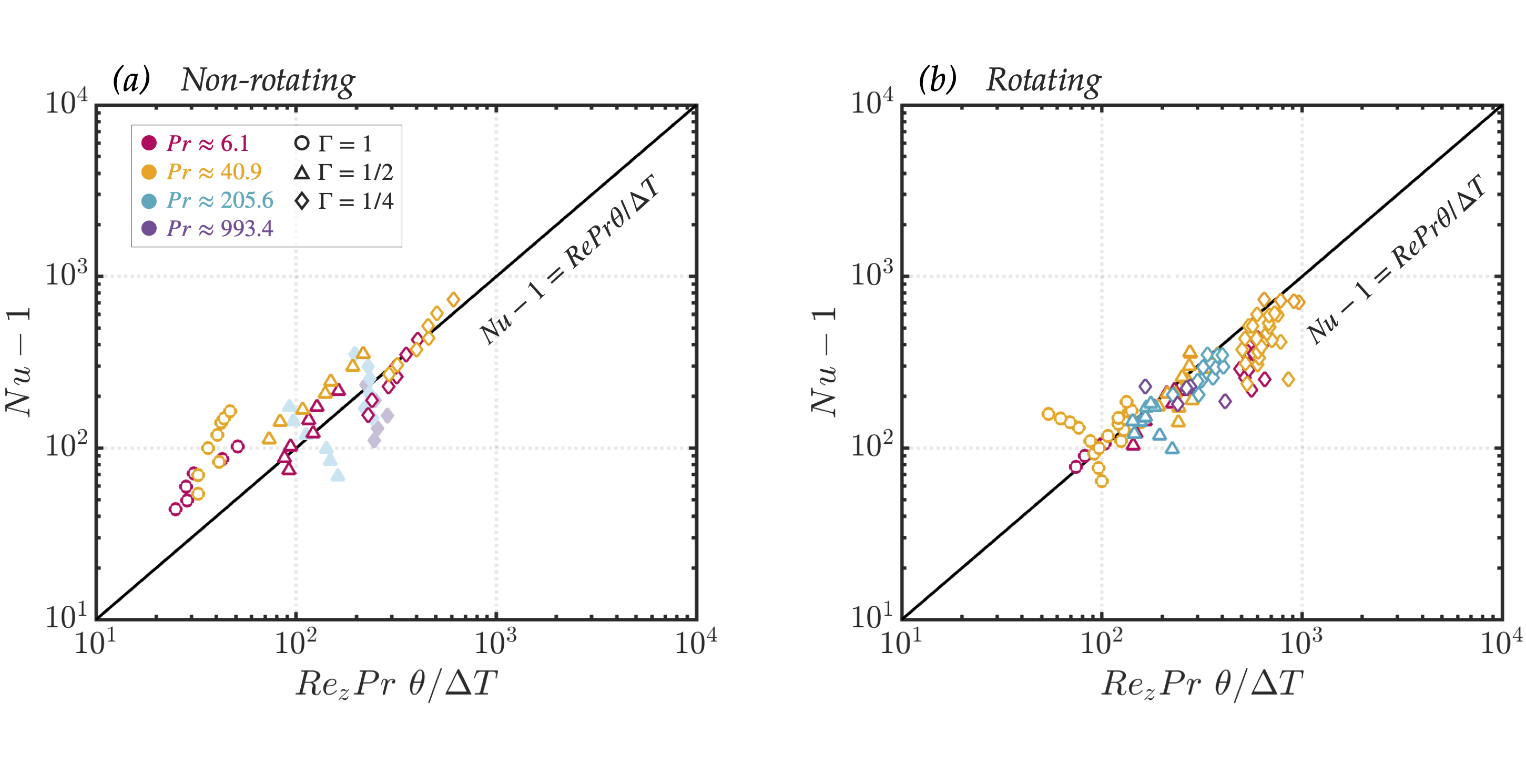}}
\caption{Correlation between $Nu-1$ and $Re_z Pr \theta / \Delta T$. Panel (a) shows non-rotating measurements, while (b) shows rotating measurements to test the accuracy of (6). Semi-transparent markers in (a) indicate measurements of $\theta$ located within the viscous boundary layer, $\delta_{\nu} = 0.25H/\sqrt{Re_z}$. (Colour  online)}  %
\end{center}
\end{figure}

Figure 5a shows these quantities for the non-rotating convection cases. The $Pr$ = 6.1 and 40.9 data trend approximately linearly, albeit with slight dependence on aspect ratio. The $Pr$ = 206 and 993 data scale more steeply, which can be explained by the location of the internal temperature measurement, $\theta$, relative to the cell boundary. All $\theta$ are measured 5.0 cm below the top boundary of the convection cell (while velocity measurements are made at the mid-height). The viscous boundary layer thickness is given by $\delta_{\nu}/H = 0.25/\sqrt{Re}$, which is larger then 5.0 cm for the higher $Pr$, low $Re_z$, non-rotating cases \cite[]{schlichting2016boundary}. This places all $Pr$ = 206 and 993 data inside the boundary layer, rather than the fluid bulk, such that $\theta$ has a much different $Ra$ dependence than the bulk flow expectation, and therefore is inconsistent with our measured $Re_z$ \cite[cf.][]{sun2008experimental}. All $Pr$ = 6.1 and 40.9 data is measured significantly past the boundary layer, and is thus better described by (6). 

Figure 5b shows the same relation, but for rotating cases. All rotating measurements of $\theta$ were located outside of the boundary layers and within the fluid bulk for all Prandtl numbers. This is reflected by the near linear trend. The slight aspect ratio dependence present in figure 5a is also no longer present.

\subsection{Non-rotating convection scalings}
Non-rotating convection experiments are carried out to benchmark the system and to provide a basis for comparatively analyzing the effect of rotation on the flow. Figure 6 summarises the non-dimensional RBC measurements. Table B2 displays the corresponding best-fit information for each measured value and fluid.

Figure 6a shows $Nu$ data plotted as a function of $Ra$, while figure 6b shows the best-fit scaling exponent to $Nu\sim Ra^{\alpha}$ versus $Pr$. The result across all $Pr$ aligns closely with the classical heat transfer scaling exponent of $\alpha=1/3$, with no clear dependence on the Prandtl number or aspect ratio. Collapsing this data in the form $Nu = cRa^{\alpha}Pr^{\beta}$ yields
\begin{equation}
    Nu = (0.075\pm0.10) \ Ra^{0.319 \pm 0.010} \ Pr^{0.047\pm0.032}.
\end{equation}
The $Ra$ and $Pr$ dependences in (47) agree well with other studies finding $ 0.289 \lesssim \alpha \lesssim 1/3 $ and $ -0.03 \lesssim \beta \lesssim 0.074 $ \cite[][]{globe1959natural, xia2002heat, li2021effects}.

\begin{figure}
\begin{center}
\resizebox{1.0\linewidth}{!}{\includegraphics{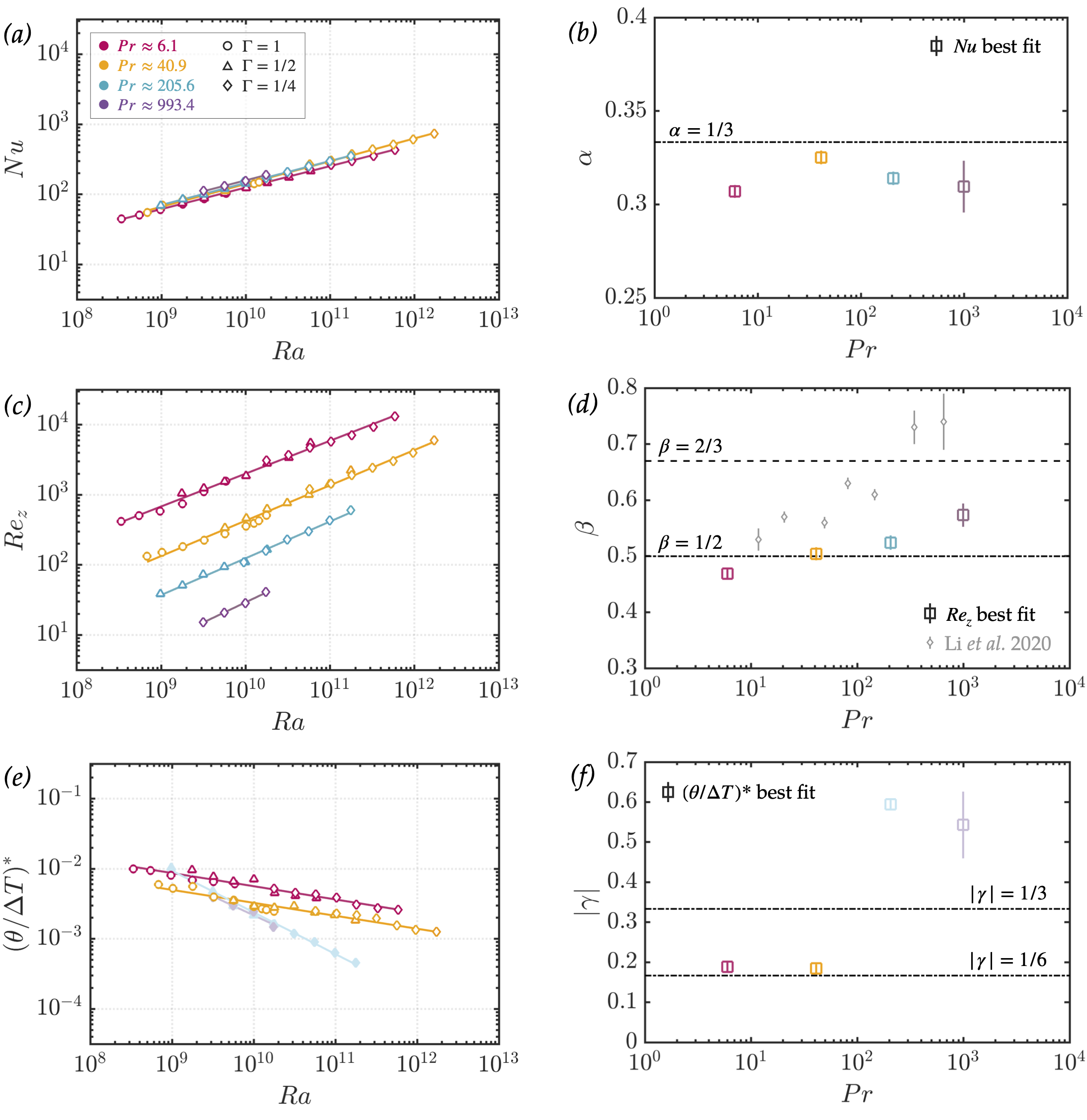}}
\caption{Non-rotating measurements. Panel (a) displays global heat transfer measurements, with panel (b) highlighting the best-fit scaling exponent, $\alpha$, against $Pr$. Panel (c) displays Reynolds number measurements, with panel (d) similarly highlighting the best-fit scaling exponent, $\beta$, against $Pr$. Cubic convection results from \cite{li2021effects} are included in (d) for comparison. Panel (e) shows measurements of the normalised internal temperature perturbation, where $(\theta / \Delta T)^* = (\theta / \Delta T) \cdot{} \varGamma^{c}$. A best-fit to the data yields $c = 0.6$ for $Pr = 6, \ 41$ (measured inside the fluid bulk) and $c = 1.0$ for $Pr = 210, \ 993$ (measured within the viscous boundary layer). The $Pr=210, \ 993$ measurements of $\theta$ cannot be compared to bulk scaling theory and are thus denoted here with semi-transparent markers. (Colour  online)}%
\end{center}
\end{figure}
 
Figure 6c shows $Re_z$ versus $Ra$, and figure 6d shows the best-fit scaling exponent to $Re_z\sim Ra^{\beta}$ versus $Pr$. The result is arguably near the inertial scaling prediction of $\beta=1/2$ across all $Pr$. However, the exponent consistently increases with Prandtl number, approaching the high $Pr$ boundary-layer controlled prediction of $\beta=2/3$. This trend is consistent with the result of \cite{li2021effects}, who performed silicone oil RBC experiments in a cubic cell, the results of which are included in figure 6d for comparison. Our largest Prandtl number data set ($Pr=993$) has not reached the $\beta=2/3$ prediction, with a best-fit yielding $\beta_{Pr=993}=0.573$. This result is comparable to direct numerical simulations performed by \cite{horn2013non}, who found $Re\sim Ra^{0.583}$ for $Pr=2548$ in a 3D cylindrical cell. The $Pr=6$ scaling, $\beta_{Pr=6}=0.469$, also compares well with prior velocity studies, all of which find scaling exponents near $\beta=1/2$ \cite[]{qiu2004velocity, daya2001does, hawkins2023laboratory}. Collapsing the $Pr\geq6$ data gives a scaling relation of 
\begin{equation}
    Re_{z} = (0.086\pm0.016)\ Ra^{0.497\pm0.031} \ Pr^{-0.797\pm0.037}.
\end{equation}
However, the variation in $Ra$ dependence across $Pr$ suggests a single collapse across this wide of a $Ra$ and $Pr$ range will not be as accurate as independent fits.
 
Figure 6e shows modified $\theta/\Delta T$ (denoted with a *) plotted versus $Ra$, while figure 6f shows the best-fit scaling exponent to $(\theta / \Delta T)^* \sim Ra^{\gamma}$ versus $Pr$. Following Li et al. (2020), the modification corrects for an aspect ratio dependence, where a best-fit to $(\theta / \Delta T)^* = (\theta / \Delta T) \cdot{} \varGamma^{c}$ yields $c=0.6$ for $Pr=(6,41)$ and $c=1.0$ for $Pr=(210,993)$. As discussed in section 4.2, the high $Pr$ non-rotating measurements of $\theta$ are sampled within the viscous boundary layer (denoted with faded markers), while the moderate $Pr$ measurements are all within the fluid bulk. Thus, we perform two separate fits to find the aspect ratio correction. The $(\theta / \Delta T)^*$ scaling behaviour for moderate $Pr$ is well-predicted by (10c), which estimates $\gamma=-1/6$. The scaling behaviour for high $Pr$ deviates significantly from the $\gamma=-1/3$ prediction of (11c), however, this result is reasonable for flow within the viscous boundary layer \cite[cf.][]{sun2008experimental}. Overall, the non-rotating RBC measurements compare well with both theoretical predictions and a variety of prior work, verifying the accuracy of this convection system and the accompanying diagnostics.

\subsection{Rotating convection scalings}

\begin{figure}
\begin{center}
\resizebox{1.0\linewidth}{!}{\includegraphics{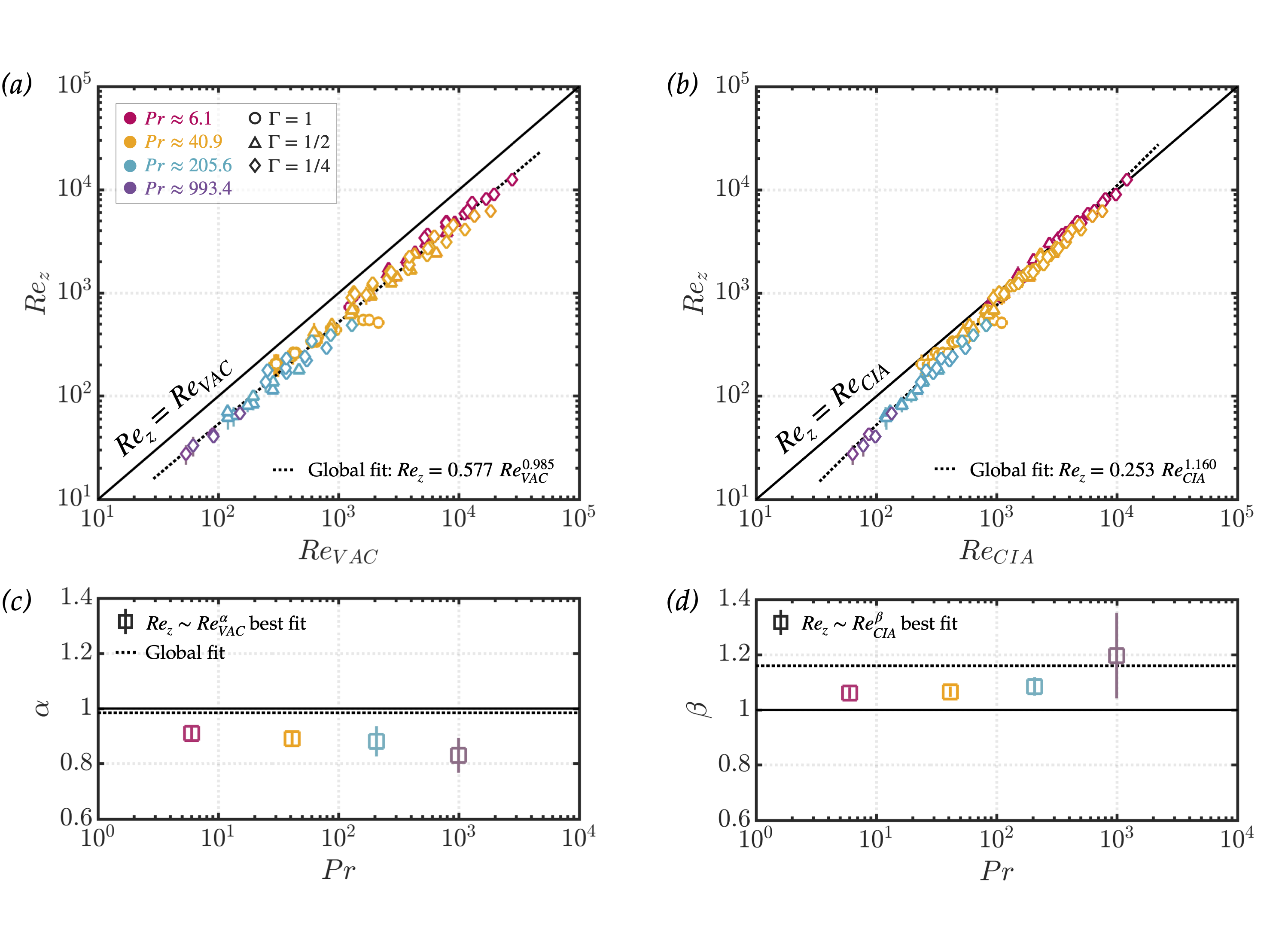}}
\caption{Rotating velocity scaling results. Panels (a) and (b) show $Re_z$ vs. $Re_{VAC}$ and $Re_z$ vs. $Re_{CIA}$, respectively, for all Prandtl numbers. Error bars represent systematic error calculated with (46). Panels (c) and (d) show $\alpha$ vs. $Pr$ for the $Re_{VAC}$ scaling and $\beta$ vs. $Pr$ for the $Re_{CIA}$ scaling, respectively. Square markers denote the best-fit scaling exponent for each value of $Pr$. Vertical error bars represent statistical error from the fit. The solid black line represents a scaling exponent of unity. The dotted black line is the best global fit to all $Pr\geq 6$. (Colour  online)}%
\end{center}
\end{figure}

Figures 7a and 7b show the measured Reynolds number versus the VAC scaling prediction (23) and CIA scaling prediction (28), respectively. Figures 7c and 7d detail the corresponding best-fit exponents, $\alpha$ and $\beta$, displayed as square markers. The solid black line marks a scaling exponent of unity and the dotted black line marks the value corresponding to a global best-fit on $Re_z$ across all values of $Pr\geq6$.

These data show scalings near unity for both VAC and CIA Reynolds number estimates. Global fits yield
\begin{equation}
    Re_{z} =(0.577 \pm 0.059) \ Re_{VAC}^{0.985 \pm 0.013},
\end{equation}
\begin{equation}
    Re_{z} =(0.253 \pm 0.020) \ Re_{CIA}^{1.160 \pm 0.011},
\end{equation}
with both $Re_{VAC}$ and $Re_{CIA}$ closely predicting velocities for the parameters covered here. Table B3 summarises the individual fits for each $Pr$ fluid. These individual $Pr$ fits have similar scaling exponents, however, the CIA coefficients do differ with $Pr$. This suggests some $Pr$-dependent shingling in the CIA framework \cite[cf.][]{cheng2016tests}.

\begin{figure}
\begin{center}
\resizebox{1.0\linewidth}{!}{\includegraphics{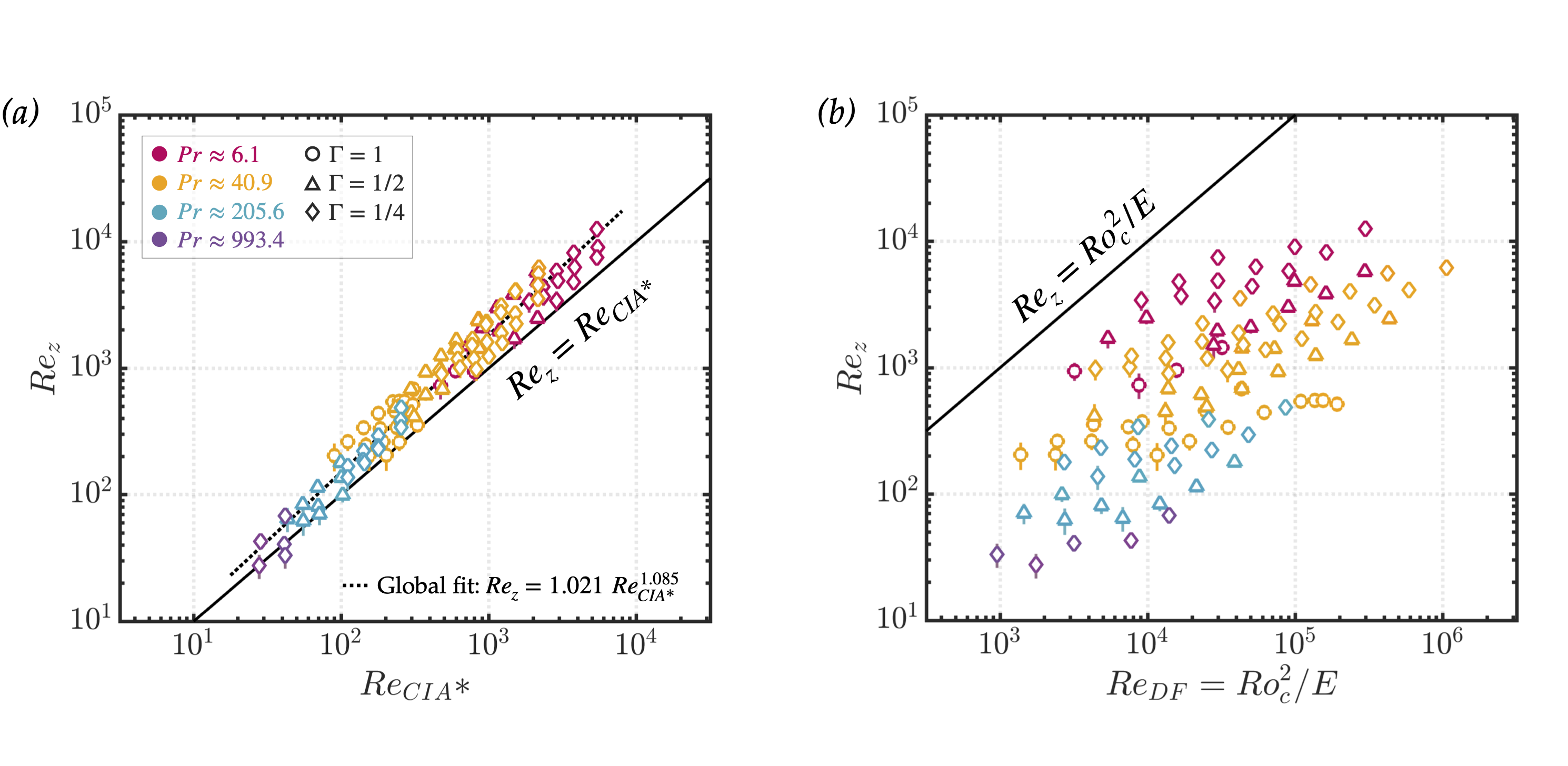}}
\caption{Tests of rotating diffusivity-free scaling estimates. Panel (a) shows measured $Re_z$ versus the CIA* prediction, which is formulated with the diffusivity-free length scale $\ell/H \sim Ro_c$ estimate but employs the measured heat transfer efficiency, $Nu$. Panel (b) shows measured $Re_z$ versus $Re_{DF}$, which substitutes $Nu$ for the asymptotic diffusivity-free heat transfer prediction. (Colour  online)}%
\end{center}
\end{figure}

Figure 8 tests the diffusivity-free (DF) scaling estimates from section 2.4.3. Figure 8a shows $Re_z$ versus the $Re_{CIA*}$ prediction from (36), which specifically tests the DF length scale $\ell_{DF}/H\sim Ro_c$, but employs the measured heat transfer efficiency, $Nu$. The data is collapsed relatively well across all Prandtl numbers, with a global scaling exponent near unity: 
\begin{equation}
    Re_{z} = (1.021 \pm 0.142) \ Re_{CIA*}^{1.085 \pm 0.021}.
\end{equation}
This implies that estimating $\ell/H\sim Ro_c$ as the convective length scale in the fluid bulk is valid for our rotating flows, even if the global heat transfer is still controlled by boundary layer diffusion. 

Figure 8b shows the measured Reynolds number versus the fully diffusivity-free Reynolds scaling, $Re_{DF} = Ro_c^2/E$. This prediction uses the same length scale estimate, $\ell_{DF}$, but also assumes an asymptotic diffusion-free heat transfer law given by (35). The data is no longer collapsed, indicating that the heat transfer in these experiments remains significantly affected by boundary layer physics \cite[cf.][]{hawkins2023laboratory}. 

\begin{figure}
\begin{center}
\resizebox{1.0\linewidth}{!}{\includegraphics{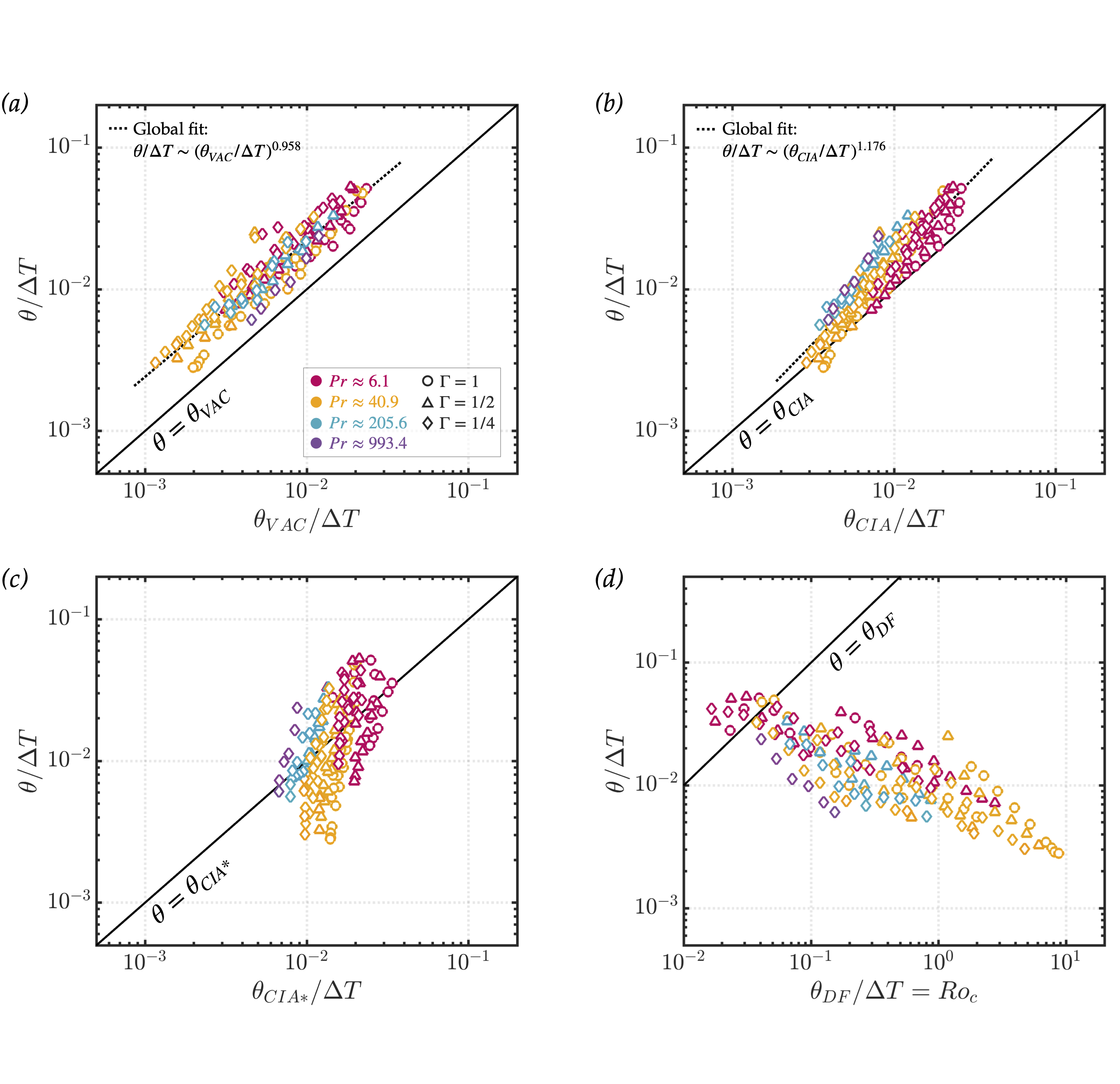}}
\caption{Tests of VAC, CIA, and diffusivity-free scaling estimates for $\theta / \Delta T$. Panels (a) and (b) show $\theta / \Delta T$ versus the corresponding VAC and CIA predictions, respectively. Panel (c) shows $\theta / \Delta T$ against the CIA* prediction, which tests the diffusivity-free length scale prediction, $\ell/H \sim Ro_c$, while using the measured $Nu$ values. Panel (d) tests the fully diffusivity-free prediction, $\theta_{DF} / \Delta T = Ro_c$, in which measured $Nu$ is substituted by the asymptotic prediction of $Nu-1\sim Ro_cRaE$ \cite[]{julien2012heat}. (Colour  online)}%
\end{center}
\end{figure}

An alternative approach considers the normalised local temperature perturbation, $\theta / \Delta T$. The VAC prediction for $\theta / \Delta T$ (24) is shown in figure 9a, and the CIA prediction (29) is shown in figure 9b. Measurements of $\theta / \Delta T$ are acquired independently of $Re_z$, thus providing an alternative bulk characteristic to test against theoretical prediction. Similar to the Reynolds number scalings, the VAC estimate provides a closer fit to the data than the CIA estimate, but both are near unity. Global fits yield
\begin{equation}
    \theta/\Delta T =(1.804 \pm 0.309) \ (\theta_{VAC}/ \Delta T)^{0.958 \pm 0.031},
\end{equation}
\begin{equation}
    \theta/\Delta T =(3.636 \pm 0.762) \ (\theta_{CIA}/ \Delta T)^{1.176 \pm 0.040}.
\end{equation}

Figure 9c shows $\theta / \Delta T$ versus the $\theta_{CIA*} / \Delta T$ prediction from (37), which tests $\ell_{DF}$ while maintaining the measured heat transfer efficiency, $Nu$, as in figure 8a. Figure 9d tests the fully diffusivity-free scaling, $\theta_{DF} / Ro_c$. This prediction uses the same length scale estimate, $\ell_{DF}$, but also assumes asymptotic heat transfer given by (35). While the scaling for $\theta_{CIA*} / \Delta T$ is better collapsed than that for $\theta_{DF} / \Delta T$, neither adequately describes the data. 

Importantly, we see in figure 9 that panels (a) and (b) collapse the data well, whereas the data does not agree that well with theory in panels (c) and (d). This indicates that $Nu$ is required to fit the local bulk heat transfer data.  Further, the $\theta/\Delta T$ data greatly differ from the diffusivity free prediction in panel (d). Thus, the local bulk heat transfer processes are still affected by diffusivity-controlled boundary layer processes in our experiments, in good agreement with the arguments of \cite{oliver2023small}.

\section{Discussion}

\subsection{Length scale considerations}
The comparable scaling results of CIA and VAC Reynolds numbers warrant analysis of the theoretical length scales that distinguish them ($\ell_{turb}$ and $\ell_{crit}$, respectively). The ratio of these two scales is given as
\begin{equation}
    \frac{\ell_{turb}}{\ell_{crit}} = 
    \begin{dcases}
    \frac{Ro^{1/2}} {2.4 E^{1/3}} & Pr > 0.68 \\
    \frac{Ro^{1/2}} {2.4 (E/Pr)^{1/3}} & Pr \leq 0.68 \ ,
    \end{dcases}
\end{equation}
where $Ro$ is the measured Rossby number. Figure 10 shows this ratio plotted against  the super-criticality, $\widetilde{Ra}$, defined as
\begin{equation}
    \widetilde{Ra} =
    \begin{dcases}
    \frac{Ra} {8.7 E^{-4/3}} & Pr > 0.68 \\
    \frac{Ra} {17.4 (E/Pr)^{-4/3}} & Pr \leq 0.68
    \end{dcases}
\end{equation}
\cite[]{chandrasekhar1961}. Liquid metal rotating convection data from \cite{vogt2021oscillatory} is included to consider a low Prandtl number fluid ($Pr=0.027$), as well as water rotating convection data from \cite{madonia2023reynolds} to consider more extreme cases ($3 \times 10^{11} \lesssim Ra \lesssim 4 \times 10^{12}$ and $E = 5 \times 10^{-8}$). Figure 10 shows that $\ell_{turb} / \ell_{crit}$ deviates from unity by less than one order of magnitude across five orders of magnitude in $\widetilde{Ra}$ and five orders of magnitude in $Pr$. Importantly, this invariability indicates that a meaningful separation of these two theoretical scales is unlikely for the range of $Ra$, $E$, and $Pr$ covered here and in the vast majority of current day studies of rotating convection and dynamo action \cite[e.g.,][]{yadav2016approaching, aurnou2017cross, guervilly2019turbulent}.

Focusing on high $Pr$ rotating fluid dynamics, our $6 \leq Pr \leq 993$ data is well collapsed by the empirical expression
\begin{equation}
    \ell_{turb} / \ell_{crit} = (2.037 \pm 0.221) \ \widetilde{Ra}{}^{0.213 \pm 0.009} \ Pr^{-0.478 \pm 0.021}.
\end{equation}
Inverting (56) for a scale separation of $\ell_{turb} / \ell_{crit} \approx 100$ in water ($Pr=6$) yields a super-criticality of $\widetilde{Ra}\approx10^9$. A laboratory experiment with $E\approx10^{-6}$ would then require a Rayleigh number of $Ra\approx10^{18}$ and convective Rossby number $Ro_c\approx10^{2}$ to see this increased scale separation. These values are not only difficult to obtain, but are also no longer relevant to rotating dynamics since $Ro_c \gg 1$. Higher $Pr$ requires even higher $Ro_c$ to separate the turbulent scale from the critical onset scale. This implies that rapid rotation, and thus extreme control parameter values, will be necessary to robustly disambiguate the onset and turbulent scales of rotating convection.

\begin{figure}[b]
\begin{center}
\resizebox{0.7\linewidth}{!}{\includegraphics{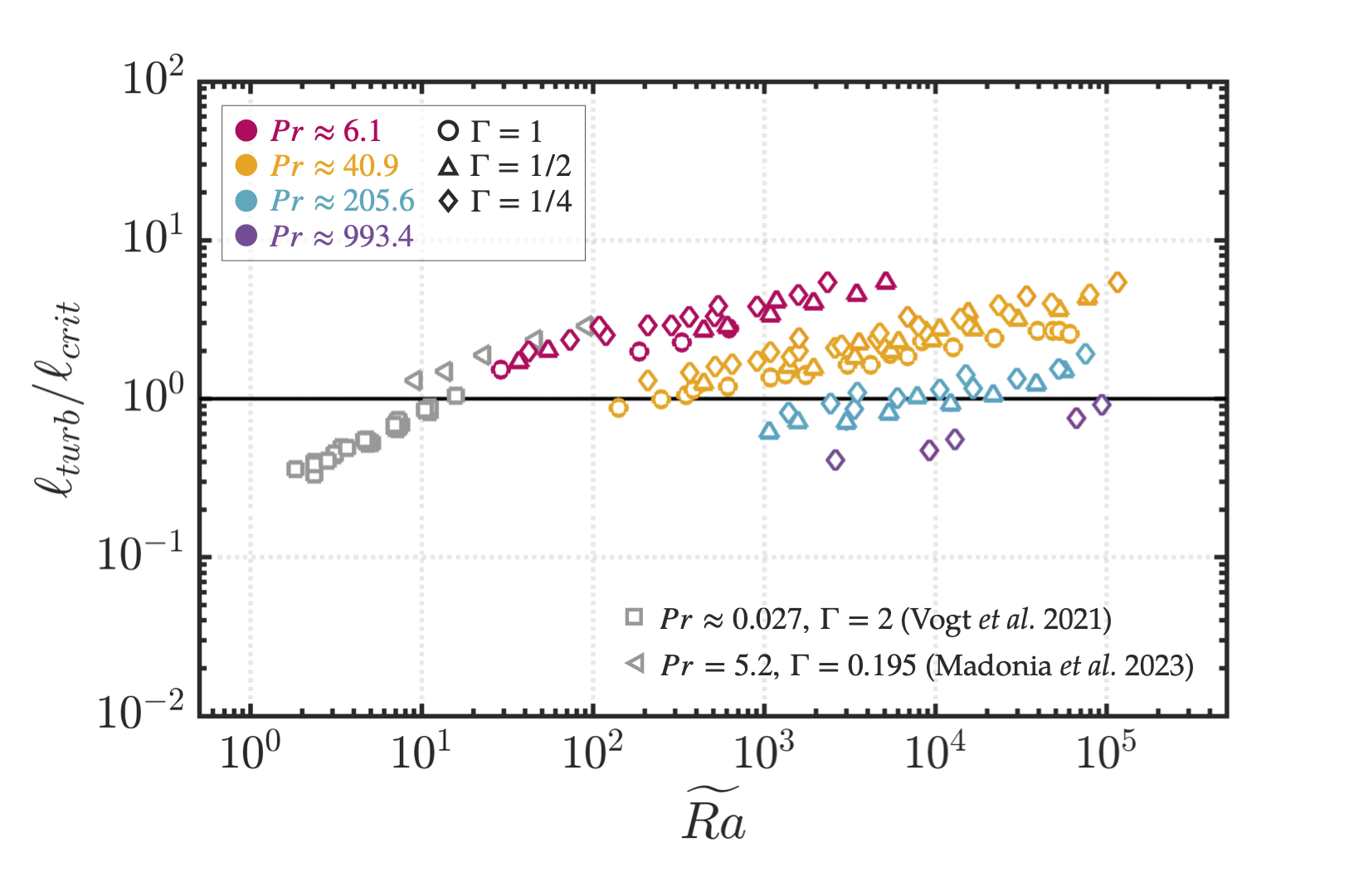}}
\caption{Comparison of convective length scale estimates, including data from \cite{vogt2021oscillatory} and \cite{madonia2023reynolds}. The ratio $\ell_{turb}/\ell_{crit}$ is within an order of magnitude of unity over a span of five orders of magnitude in super-criticality, $\widetilde{Ra}$, indicating a lack of scale-separability. (Colour  online)}%
\end{center} 
\end{figure}

In Earth's core, an analysis of secular variation yields a Rossby number estimate of $Ro = ReE \simeq 4 \times 10^{-6}$ \cite[]{jackson2015geomagnetic}. The corresponding Reynolds number is $Re \simeq 4 \times 10^9$ assuming $E \approx 10^{-15}$. If the turbulent $Pr = 1$ estimate is used, then equation (57) in section 5.2 predicts a Rayleigh number of $Ra\simeq 7 \times 10^{23}$ and (56) predicts $\ell_{turb} / \ell_{crit} \simeq 8$. If the compositional $Pr = 100$ estimate is used, then (57) predicts $Ra\simeq 2 \times 10^{27}$ and (56) predicts $\ell_{turb} / \ell_{crit} \simeq 5$. Importantly, both of these low convective Rossby estimates ($Ro_c \simeq 5 \times 10^{-3}$ and $Ro_c \simeq 8 \times 10^{-4}$, respectively) yield a turbulent length scale estimate comparable to the onset scale.

\subsection{Global collapse of Reynolds number measurements}
The theoretical VAC and CIA balances explored here both adequately describe our data. Nevertheless, we explore an empirically-determined predictive scaling for the Reynolds number that robustly describes convective $u_z$ velocities present in rotating RBC systems. 

We first perform a non-linear fit to the Reynolds number measurements for scaling dependence on the Rayleigh, Ekman, and Prandtl numbers, which were used as control parameters in the experiments. This provides a $\Delta T$-based predictive scaling of the form $Re = c Ra^{\alpha} E^{\beta} Pr^{\gamma}$:
\begin{equation}
    Re_{\Delta T}=(0.241 \pm 0.034) Ra^{0.563 \pm 0.010}E^{0.213 \pm 0.011}Pr^{-0.944\pm0.010}.
\end{equation}
The corresponding dimensional $u_z$ velocity scaling is
\begin{equation}
    u_{\Delta T} = (0.208 \pm 0.056) (\alpha g \Delta T)^{0.563 \pm 0.034} H^{0.263 \pm 0.021} \varOmega^{-0.213\pm0.011} \nu^{-0.294 \pm 0.031} \kappa^{0.381 \pm 0.020}.
\end{equation}
We additionally consider the measured Nusselt number as a predictive parameter for the resulting velocity. The results presented in section 4.4 demonstrate the importance of considering heat transfer behaviour when examining velocity scaling behaviour. The Rayleigh number is thus replaced with a flux-Rayleigh number defined as 
\begin{equation}
    Ra_F = RaNu = \frac{\alpha g q H^4}{\kappa \nu k}.
\end{equation}
This provides a $q$-based predictive scaling of the form $Re = c Ra_F^{\alpha} E^{\beta} Pr^{\gamma}$:
\begin{equation}
    Re_{q}=(0.678 \pm 0.082) Ra_F^{0.435 \pm 0.006}E^{0.228 \pm 0.009}Pr^{-0.966\pm0.009} ,
\end{equation}
with a corresponding dimensional scaling of
\begin{equation}
    u_{q} = (0.579 \pm 0.100) (\alpha g q / k)^{0.435 \pm 0.006} H^{0.284 \pm 0.042} \varOmega^{-0.228\pm0.009} \nu^{-0.173 \pm 0.024} \kappa^{0.531 \pm 0.015} .
\end{equation}

The empirical fits presented in (57-61) are determined from data spanning a wide range of $Ra$, $E$, and $Pr$, and may therefore be useful in describing a variety of rotating RBC systems. Note that our data predominantly follows a $Nu\sim Ra^{1/3}$ heat transfer scaling (see figure A1a), which should be considered when utilising these empirical findings \cite[]{cheng2015laboratory, aurnou2020connections, cheng2020laboratory, oliver2023small, hawkins2023laboratory}. Further, our measured $Ro_z$ values range from $4\times10^{-4}$ to $2\times10^{-1}$, corresponding to a convective Rossby range of $10^{-2} \lesssim Ro_c \lesssim 10^1$. In contrast, we take $Ro \sim 4 \times 10^{-6}$ in Earth's core \cite[]{jackson2015geomagnetic} and estimate $Ro_c = \sqrt{Ra E^2/Pr} \sim 3 \times 10^{-3}$ for $E \simeq 10^{-15}$, $Ra \simeq 10^{24}$ and $Pr \sim 0.1$ \cite[]{cheng2016tests}.

\subsection{Velocimetry error and sidewall effects}
Figure 11a shows all measured velocities versus $u_{\Delta T}$ given by (58). In the results shown thus far, velocity measurements encumbered by a large noise-to-signal ratio have been removed per the methods discussed in section 3.4. These values are included in figure 11 to demonstrate the additive effect of system noise on the LDV data. The data tail off at approximately the same dimensional velocity, corresponding to the minimum measurable rms-velocity discussed in section 3.4. Figure 11b shows the same data, but plotted non-dimensionally against $Re_{\Delta T}$ given by (57). Shallow tails emerge from the otherwise linear trend for each $Pr$--$E$ shingle.

\begin{figure}
\begin{center}
\resizebox{1.0\linewidth}{!}{\includegraphics{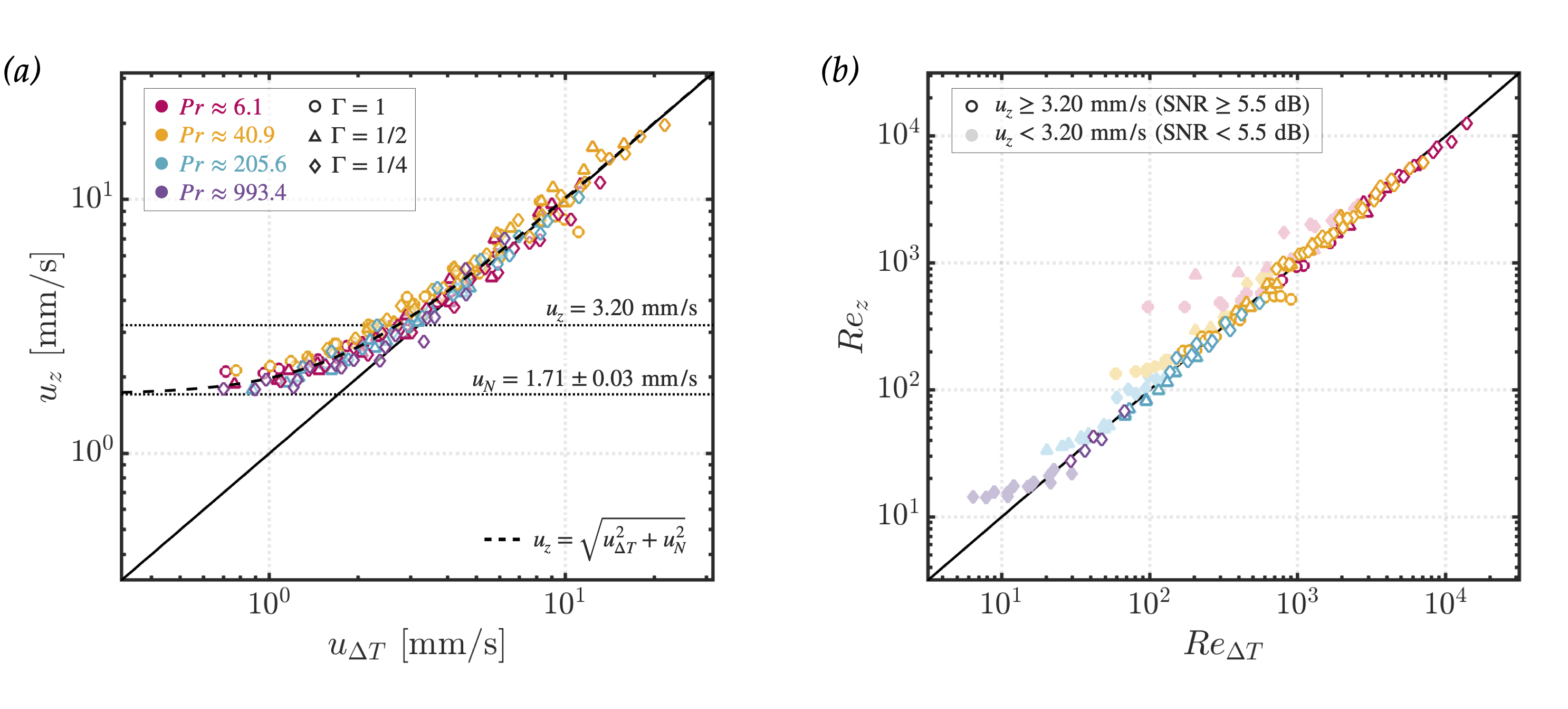}}
\caption{Error model analysis. a) All dimensional velocity measurements, $u_z$, versus the empirical prediction given by (58), which was determined exclusively from SNR $\geq$ 5.5 dB measurements. The upper dotted line indicates the cut-off value for SNR = 5.5 dB (below which has been removed from the analysis thus far). The lower dotted line represents the best-fit noise floor to (62b), $u_N = 1.71$ mm/s. The bold dashed line shows (62b) as the model equation. b) All Reynolds number measurements, $Re_z$, against the empirical prediction given by (57). Boldly coloured markers are high SNR points that have been used in the analysis thus far. Semi-transparent markers are the low SNR points which fall below the dotted $u_z=3.20$ mm/s line in panel (a). (Colour  online)}%
\end{center}
\end{figure}

The velocity measured by the LDV contains the actual fluid motion signal convoluted with system noise, both of which maintain an independent zero-mean Gaussian-distributed profile (see figure 3c). The resulting velocity distribution can be modeled as a convolution of two Gaussian PDFs, population $f$ and $g$. The convolved distribution is also Gaussian and is described by 
\begin{subequations}
\begin{gather}
    \mu_{f \circledast g} = \mu_f + \mu_g, \qquad 
    \sigma_{f \circledast g} = \sqrt{\sigma_{f}^2 + \sigma_g^2},
    \tag{\theequation a-b}
\end{gather}
\end{subequations}
where $\mu$ is the mean and $\sigma$ is the standard deviation \cite[]{bromiley2003products}. The time-averaged mean for all cases presented here is approximately zero, so the rms-velocities are treated as equivalent to the standard deviation. Equation (62b) is plotted as a dashed line in figure 11a, where $\sigma_f$ and $\sigma_g$ are replaced with $u_{\Delta T}$ (representing the flow signal rms) and $u_{noise}$ (representing the noise rms), respectively. A fit of the velocimetry data to (62b) yields $u_{noise} = 1.71 \pm 0.03$ mm/s as the best-fit noise floor. This value agrees well with the experimentally-determined noise level of 1.69 mm/s found by linear spin-up tests (see section 3.4). The agreement in trend between the velocity data and error model presented in figure 11a, as well as the isothermal spin-up experiment result, supports system noise as the origin of the deviating tails in the LDV data.

It is important to consider the effect of the solid sidewall on the interior bulk flow dynamics in geometrically confined flows. Several studies have characterised the sidewall circulation (or boundary zonal flow) as a region of either cyclonic or anti-cyclonic flow at the cell boundary with enhanced heat transport relative to the bulk flow \cite[]{horn2017prograde, favier2020robust, de2020turbulent, zhang2021boundary, lu2021heat, grannan2022experimental}. These so-called `wall modes' have further been suggested to generate jets that emanate from the sidewall circulation into the fluid bulk.  These jets could enhance otherwise low bulk flow velocities, particularly those in the columnar convection regime \cite[]{kunnen2021geostrophic, madonia2021velocimetry, madonia2023reynolds}. This effect could explain the gradual plateau of the velocity data shown in figure 11a. However, the plateauing observed here occurs independent of flow regime and is well predicted by the Gaussian noise model (62b), suggesting that system noise is the more probable explanation. While it is likely that our system contains sidewall dynamics, there is no evidence that our LDV can resolve their effect due to the noise inherent to the system.

\section{Summary and future directions}
This study presents simultaneously acquired laboratory measurements of temperature, heat transfer, and axial velocity in both rotating and non-rotating convection experiments. Across 240 cases, we found robust scaling behaviours covering over four orders of magnitude in $Ra$ ($10^8 \lesssim Ra \lesssim 10^{12}$), three orders in $E$ ($10^{-4} \gtrsim E \gtrsim 10^{-7}$), and three orders in $Pr$ ($6 \lesssim Pr \lesssim 10^{3}$).  

The non-rotating results agree well with theoretical predictions and past studies for both $Nu$ and $Re$ scalings \cite[]{ahlers2009heat, li2021effects}. In rotating cases, the heat transfer predominantly follows the boundary-controlled $Nu\sim Ra^{1/3}$ scaling. The bulk convective velocities are well described by both the VAC and CIA predictions, which include the measured heat transfer, $Nu$, as an input parameter. We additionally test a modified CIA* scaling in which the turbulent length scale $\ell/H\sim Ro^{1/2}$ is replaced with the diffusivity-free scale $\ell/H\sim Ro_c$, while the measured $Nu$ is maintained. This modified scaling did not collapse the measured $Re$ as completely, but provided an adequate prediction. Lastly, we test the fully diffusivity-free (DF) case, in which $Nu$ in the CIA* scaling is replaced with the asymptotic prediction for diffusion-free heat transfer. This DF scaling did not collapse our data. Overall, the predictions derived from a local baroclinic torque balance provided sufficient estimates of bulk convective velocities, but only when the measured heat transfer was accounted for in the prediction. Our results show that the bulk interior flows can be described by inviscid dynamics. However, heat transfer is consistently controlled by the boundary layers and thus has yet to reach the asymptotic diffusivity-free regime \cite[cf.][]{bouillaut2021experimental, kolhey2022influence, oliver2023small, hawkins2023laboratory}.

The agreement amongst both the VAC and CIA velocity scale predictions implies that the onset and turbulent length scales have not yet achieved a strong scale separation in our experimental range (figure 10). We find that a large super-criticality, $\widetilde{Ra}$, is required to yield a significant difference between the two scales. Achieving this in a low $Ro\ll1$ regime relevant to geostrophic flows would necessitate very extreme $Ra$ and $E$, such that even Earth's core may have comparable onset and turbulent length scales.

Further theoretical and experimental investigation of the characteristic length scales of rotating convection systems is crucial to advance our understanding of what appears to be the simultaneous VAC and CIA balance present in our experiments. Global heat transfer efficiency, $Nu$, and local cross-axial length scale, $\ell$, are two parameters that our results have shown to be essential to these balances, motivating future studies with simultaneous heat transfer and velocity field measurements. 

Our laboratory experimental results yield scaling laws, (57) - (61), describing rotating convection velocities across a broad range of $Pr \gtrsim 1$ fluids. These empirical scalings can be broadly applied across a range of geophysical systems, from subsurface lakes and oceans to magma oceans, as discussed in section 1. Uniting our current results with exact length scale measurements and clarifying theory will help better constrain the turbulent rotating convective flows occurring across a broad range of geophysical and astrophysical settings. 

\section*{Acknowledgements}
We thank two anonymous reviewers, whose feedback substantively improved this manuscript, and K. Julien for numerous fruitful discussions. This research was supported by the National Science Foundation (J.M.A., EAR 1620649; EAR 1853196) and the National Defense Science and Engineering Graduate Fellowship Program (J.A.A.).

\bibliographystyle{gGAF} 
\bibliography{PrPaper}
\markboth{J.A.~Abbate \& J.M.~Aurnou}{Geophysical \& Astrophysical Fluid Dynamics}

\appendices

\section{Rotating convection measurements}

\begin{figure}[b]
\begin{center}
\resizebox{1.0\linewidth}{!}{\includegraphics{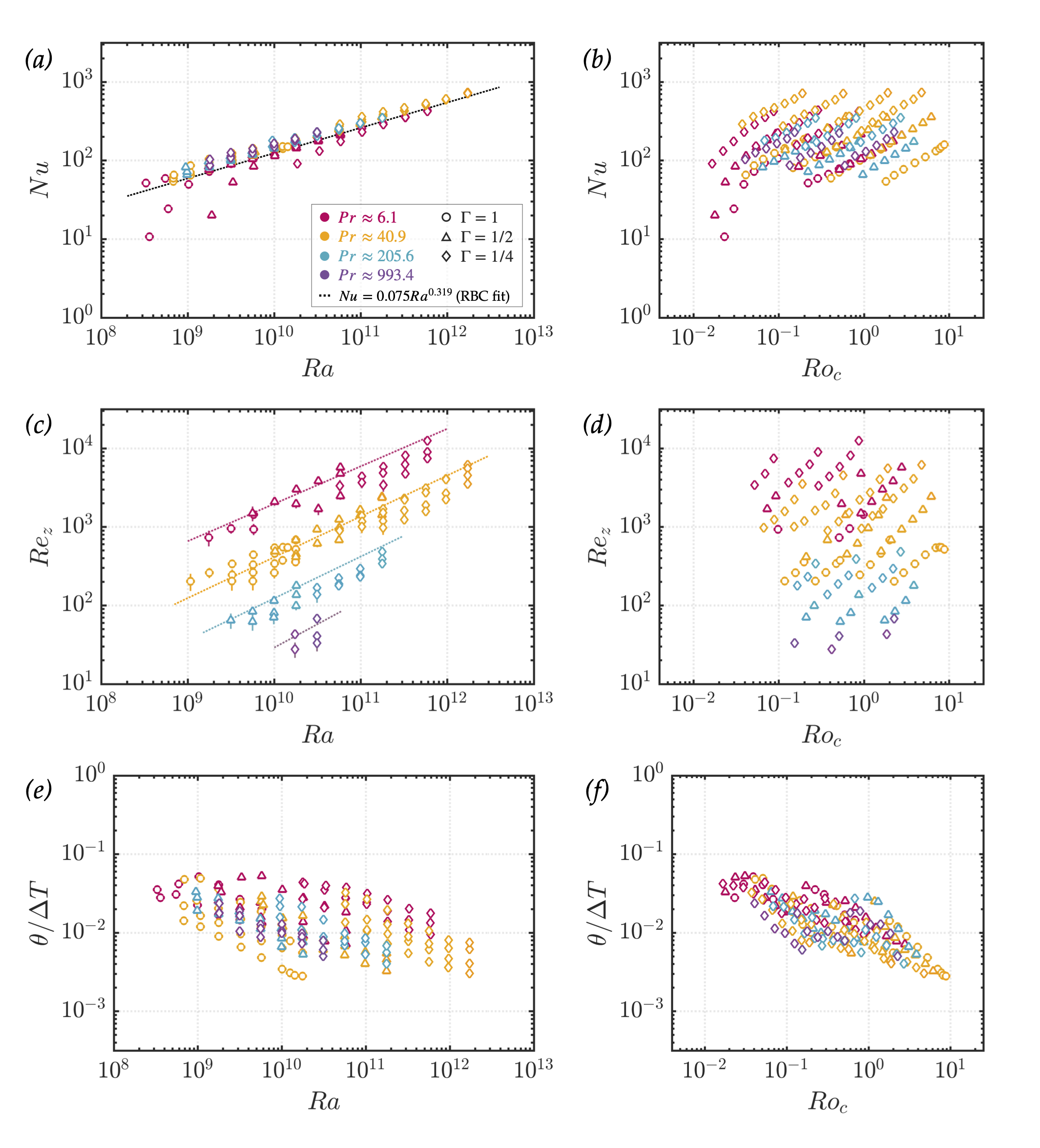}}
\caption{Rotating RBC measurements. Panels (a) and (b) show heat transfer efficiency, $Nu$, against $Ra$ and $Ro_c$, respectively. Panels (c) and (d) show the measured Reynolds number, while panels (e) and (f) show $\theta /\Delta T$, against the same quantities. (Colour  online)}%
\end{center}
\end{figure}

Figure A1 displays the non-dimensional measurements for rotating cases. Panels (a), (c), and (e) display the same quantities as figure 6 ($Nu$, $Re_z$, and $\theta / \Delta T$) for direct comparison to non-rotating measurements. Panels (b), (d), and (f) show those quantities against the convective Rossby number, $Ro_c$.

In figure A1a, $Nu$ versus $Ra$ trends are consistent with those of prior studies for $Pr=6$ rotating convection in water. For $Pr>6$, measurements largely follow the non-rotating trend, with no clear transitions in scaling exponent. This is consistent with the transition prediction given in \cite{julien2012heat}, $Ra_T \approx E^{-8/5} Pr^{3/5}$, as the range of $Ra$, $E$, and $Pr$ observed for the silicone fluids is past the predicted transition. In figure A1b, there is an approximate transition in the $Pr=6$ data at $Ro_c\approx0.6$, which is consistent with predictions for the transition to rotationally constrained flow in moderate $Pr$ fluids \cite[]{king2012heat, ecke2014heat, horn2014rotating, kunnen2021geostrophic}.

In figure A1c, $Re_z$ versus $Ra$ measurements hover around the non-rotating trend for each $Pr\geq6$ fluid, consistent with the larger values of convective Rossby that are covered.

Figure A1e shows $\theta / \Delta T$ plotted against $Ra$. Unlike the non-rotating results shown in figure 6, there is no aspect ratio-based trend to adjust for. Rather, these data are shingled by $E$ (and $Pr$). Interestingly, we see in figure A1f that even though measurements are taken at $Ro_c$ as low as 0.02 and as high as 10, there are no distinct transitions. 

Figure A2 shows the Reynolds number against the Ekman number to demonstrate the sole effect of rotation.

\begin{figure}
\begin{center}
\resizebox{0.7\linewidth}{!}{\includegraphics{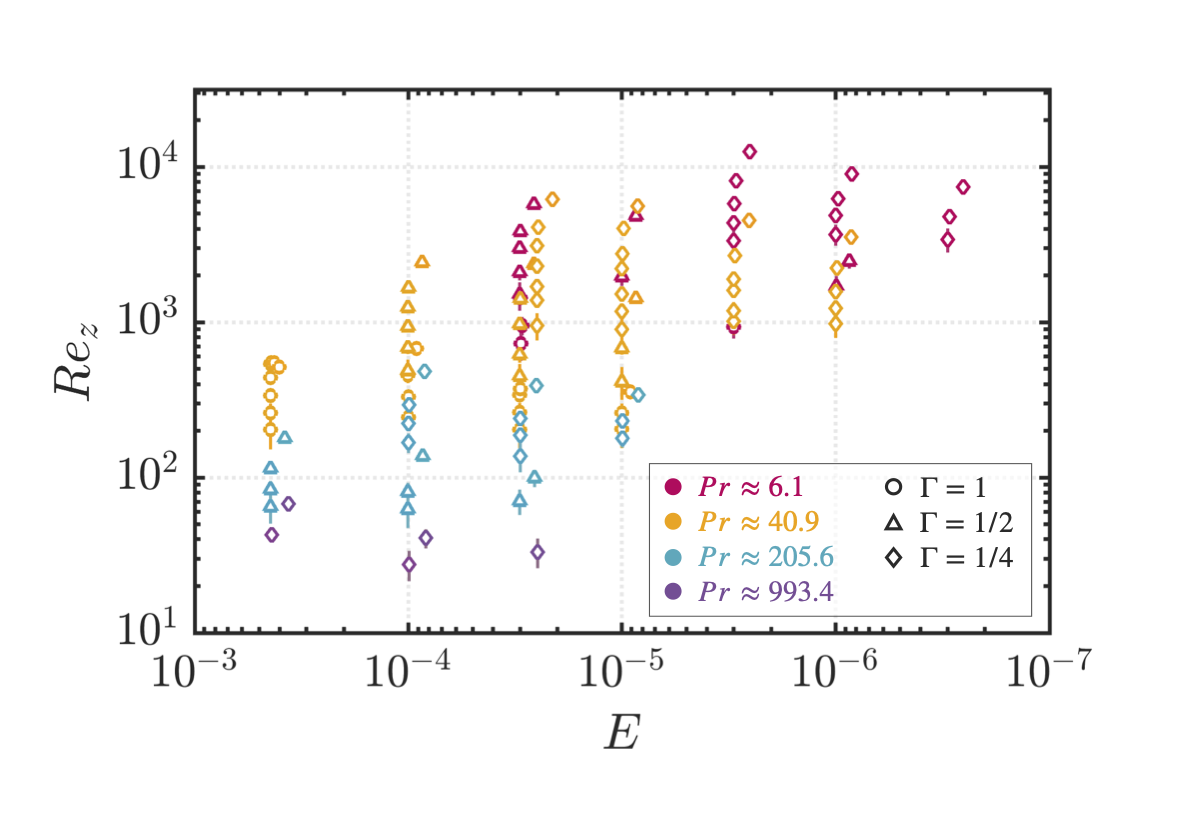}}
\caption{Reynolds number versus Ekman number for rotating RBC measurements. (Colour  online)}%
\end{center}
\end{figure}

\newpage

\section{Data Tables}

\begin{table}[h!]
\tbl{Parameters for the visualisation cases in figure 4.}
{\begin{tabular}{cccccccc}
\hline \hline
$Pr$ & $Ra$     & $E$      & $Ro_c$   & $Ra/Ra_c$ & $Re_z$   & $Ro_z$   & $Ro_z^{1/2}/(2.4E^{1/3})$  \\ 
\hline
34.5 & 1.72E+10 & $\infty$ & $\infty$ & 3.31E+06  & 5.80E+02 & $\infty$ & $\infty$  \\
34.5 & 1.72E+10 & 1.00E-04 & 2.23     & 9.18E+03  & 4.25E+02 & 4.25E-02 & 1.85      \\
34.5 & 1.72E+10 & 1.00E-05 & 0.22     & 4.26E+02  & 3.85E+02 & 3.85E-03 & 1.20      \\
172  & 3.10E+09 & $\infty$ & $\infty$ & 5.96E+05  & 6.50E+01 & $\infty$ & $\infty$  \\
172  & 3.15E+09 & 4.26E-04 & 1.82     & 1.16E+04  & 5.40E+01 & 2.30E-02 & 0.84      \\
206  & 9.98E+08 & 1.00E-04 & 0.24     & 5.32E+02  & 2.20E+01 & 2.20E-03 & 0.42      \\
814  & 5.60E+08 & $\infty$ & $\infty$ & 1.08E+05  & 5.10E+00 & $\infty$ & $\infty$  \\
814  & 5.59E+08 & 3.00E-03 & 2.49     & 2.78E+04  & 3.70E+00 & 1.11E-02 & 0.30      \\
993  & 3.14E+08 & 4.43E-04 & 0.25     & 1.22E+03  & 2.10E+00 & 9.30E-04 & 0.17      \\
\hline \hline
\end{tabular}}
\tabnote{The three independent silicone oils are separated in the table by horizontal dashed lines and are commonly referred to as $Pr$ = 41, 206, and 993 throughout the text. The Prandtl number for each fluid is not always identical because the mean temperature increases for cases with higher $\Delta T$. The Rossby number, $Ro_z$, is not directly measured here, but rather calculated from the best fit equations for each fluid presented in table B3. For rotating cases, $Ra_c$ is calculated as $Ra_c=8.7E^{-4/3}$. For non-rotating cases, we use $Ra_c$ = 5200 as determined by \cite{shishkina2021rayleigh} for a laterally insulated, no-slip, $\varGamma=1$ cylinder with isothermal top and bottom boundaries.}
\end{table}

\begin{table}[h!]
\tbl{Power-law fits for non-rotating data.}
{\begin{tabular}{ccccccc}
\hline \hline
& \multicolumn{2}{c}{$Nu = C_1Ra^{\alpha}$} & \multicolumn{2}{c}{$Re_z = C_2Ra^{\beta}$} & \multicolumn{2}{c}{$(\theta/\Delta T)* = C_3Ra^{\gamma}$} \\ 
\hline
$Pr$    & $C_1$         & $\alpha$                 & $C_2$           & $\beta$                & $C_3$            & $\gamma$                   \\
6.1   & 0.106      & 0.307$\pm$0.003     & 4.10E-02     & 0.469$\pm$0.011    & 0.435         & -0.189$\pm$0.011      \\
40.9  & 0.079      & 0.325$\pm$0.004     & 3.87E-03     & 0.504$\pm$0.012    & 0.230         & -0.185$\pm$0.010      \\
205.6 & 0.106      & 0.314$\pm$0.004     & 7.10E-04     & 0.524$\pm$0.013    & 2.11E+03      & -0.595$\pm$0.014      \\
993.4 & 0.127      & 0.310$\pm$0.014     & 5.40E-05     & 0.573$\pm$0.021    & 5.83E+02      & -0.543$\pm$0.083      \\ 
\hline \hline
\end{tabular}}
\tabnote{Individual $Pr$ best-fit power law scalings for $Nu$, $Re$, and $\theta/\Delta T$. The coefficients are $C_1$, $C_2$, and $C_3$, with exponents $\alpha$, $\beta$, and $\gamma$.}
\end{table}

\begin{table}[h!]
\tbl{Power-law fits for rotating data.}
{\begin{tabular}{ccccccccc}
\hline \hline
& \multicolumn{2}{c}{$Re_z = C_1Re_{VAC}^{\alpha}$} & \multicolumn{2}{c}{$Re_z = C_2Re_{CIA}^{\beta}$} & \multicolumn{2}{c}{$\theta/\Delta T = C_3(\theta_{VAC}/\Delta T)^{\gamma}$} & \multicolumn{2}{c}{$\theta/\Delta T = C_4(\theta_{CIA}/\Delta T)^{\zeta}$} \\ 
\hline
$Pr$    & $C_1$           & $\alpha$                  & $C_2$           & $\beta$                   & $C_3$             & $\gamma$                      &        $C_4$         &          $\zeta$             \\
6.1   & 1.220        & 0.910$\pm$0.026       & 0.591        & 1.061$\pm$0.021       & 2.944          & 1.056$\pm$0.048          & 7.161           & 1.366$\pm$0.090         \\
40.9  & 1.172        & 0.891$\pm$0.025       & 0.505        & 1.065$\pm$0.019       & 1.665          & 0.947$\pm$0.049          & 12.742          & 1.436$\pm$0.056         \\
205.6 & 0.957        & 0.881$\pm$0.054       & 0.352        & 1.085$\pm$0.034       & 2.082          & 0.989$\pm$0.056          & 18.419          & 1.436$\pm$0.061         \\
993.4 & 1.023        & 0.831$\pm$0.063       & 0.188        & 1.196$\pm$0.155       & 8.764          & 1.352$\pm$0.094          & 141.352         & 1.811$\pm$0.098         \\ 
\hline \hline
\end{tabular}}
\tabnote{Individual $Pr$ best-fit power law scalings for VAC and CIA predictions of $Re$ and $\theta/\Delta T$. The coefficients are $C_1$, $C_2$, $C_3$, and $C_4$ with exponents $\alpha$, $\beta$, $\gamma$, and $\zeta$. The scalings are generally similar across $Pr$ in both coefficient and exponent, but notably there is a $Pr$-dependent coefficient for the CIA scalings of approximately $C \sim Pr^{-0.15}$.}
\end{table}

\begin{center}
\begin{table}[h!]
\tbl{Non-dimensional, non-rotating RBC data.}
{\begin{tabular}{r@{\hspace{91mm}}}
\\
\end{tabular}}
\vspace{-11mm}
\end{table}
\tablefirsthead{%
\hline
\hline
\multicolumn{1}{c}{$\varGamma$} &
\multicolumn{1}{c}{$Pr$} &
\multicolumn{1}{c}{$Ra$} &
\multicolumn{1}{c}{$Nu$} &
\multicolumn{1}{c}{$Re_z$} &
\multicolumn{1}{c}{$\theta/\Delta T$}\\
\hline }
\tablehead{%
\multicolumn{6}{l}{\footnotesize\sl continued from previous page}\\
\hline
\multicolumn{1}{c}{$\varGamma$} &
\multicolumn{1}{c}{$Pr$} &
\multicolumn{1}{c}{$Ra$} &
\multicolumn{1}{c}{$Nu$} &
\multicolumn{1}{c}{$Re_z$} &
\multicolumn{1}{c}{$\theta/\Delta T$}\\
\hline}
\tabletail{
\hline
\multicolumn{6}{r}{\footnotesize\sl continued on next page}\\
}
\tablelasttail{
\hline
}
\begin{supertabular}{cccccc}
1.01  & 6.05 & 3.34E+08 & 44.8 & 4.19E+02 & 9.88E-03     \\
1.01  & 6.03 & 5.45E+08 & 50.7 & 5.05E+02 & 9.35E-03     \\
1.01  & 6.01 & 9.69E+08 & 60.4 & 5.86E+02 & 7.99E-03     \\
1.01  & 6.00 & 1.78E+09 & 72.1 & 7.46E+02 & 6.89E-03     \\
1.01  & 5.94 & 3.21E+09 & 86.9 & 1.11E+03 & 6.49E-03     \\
1.01  & 5.44 & 5.84E+09 & 103  & 1.56E+03 & 6.01E-03     \\
\hdashline
0.51  & 6.05 & 1.75E+09 & 75.1 & 1.05E+03 & 1.45E-02     \\
0.51  & 6.06 & 3.20E+09 & 88.7 & 1.24E+03 & 1.16E-02     \\
0.51  & 6.06 & 5.64E+09 & 103  & 1.55E+03 & 9.99E-03     \\
0.51  & 6.05 & 1.01E+10 & 123  & 1.88E+03 & 1.07E-02     \\
0.51  & 6.04 & 1.80E+10 & 146  & 2.81E+03 & 6.84E-03     \\
0.51  & 6.01 & 3.25E+10 & 175  & 3.39E+03 & 6.24E-03     \\
0.51  & 5.08 & 5.86E+10 & 217  & 5.51E+03 & 5.79E-03     \\
\hdashline
0.24  & 6.06 & 1.77E+10 & 156  & 3.09E+03 & 1.23E-02     \\
0.24  & 6.07 & 3.22E+10 & 191  & 3.70E+03 & 1.06E-02     \\
0.24  & 6.07 & 5.76E+10 & 229  & 4.71E+03 & 1.01E-02     \\
0.24  & 6.07 & 1.03E+11 & 262  & 5.78E+03 & 9.08E-03     \\
0.24  & 6.05 & 1.81E+11 & 297  & 7.09E+03 & 7.27E-03     \\
0.24  & 5.91 & 3.28E+11 & 352  & 9.28E+03 & 6.44E-03     \\
0.24  & 5.00 & 5.86E+11 & 429  & 1.32E+04 & 6.15E-03     \\
\hline
1.01  & 40.9 & 6.80E+08 & 55   & 1.34E+02 & 5.91E-03     \\
1.01  & 41.0 & 1.02E+09 & 70.5 & 1.51E+02 & 5.23E-03     \\
1.01  & 40.9 & 1.79E+09 & 83.9 & 1.82E+02 & 5.55E-03     \\
1.01  & 41.0 & 3.23E+09 & 101  & 2.25E+02 & 3.93E-03     \\
1.01  & 41.0 & 5.67E+09 & 120  & 2.76E+02 & 3.56E-03     \\
1.01  & 41.1 & 1.01E+10 & 141  & 3.56E+02 & 2.89E-03     \\
1.01  & 40.7 & 1.25E+10 & 142  & 3.92E+02 & 2.66E-03     \\
1.01  & 39.4 & 1.43E+10 & 150  & 4.27E+02 & 2.58E-03     \\
1.01  & 37.3 & 1.77E+10 & 165  & 5.11E+02 & 2.46E-03     \\
\hdashline
0.51  & 41.0 & 5.66E+09 & 113  & 3.38E+02 & 5.29E-03     \\
0.51  & 41.0 & 1.02E+10 & 143  & 4.63E+02 & 4.38E-03     \\
0.51  & 41.0 & 1.79E+10 & 169  & 6.26E+02 & 4.19E-03     \\
0.51  & 40.9 & 3.11E+10 & 209  & 7.69E+02 & 4.44E-03     \\
0.51  & 41.0 & 5.62E+10 & 246  & 1.01E+03 & 3.60E-03     \\
0.51  & 40.8 & 9.81E+10 & 299  & 1.42E+03 & 3.30E-03     \\
0.51  & 35.0 & 1.76E+11 & 355  & 2.24E+03 & 2.76E-03     \\
\hdashline
0.24  & 41.0 & 5.71E+10 & 269  & 1.20E+03 & 5.88E-03     \\
0.24  & 41.0 & 1.03E+11 & 305  & 1.44E+03 & 5.41E-03     \\
0.24  & 41.0 & 1.81E+11 & 376  & 1.91E+03 & 5.12E-03     \\
0.24  & 41.0 & 3.18E+11 & 440  & 2.43E+03 & 4.61E-03     \\
0.24  & 41.0 & 5.63E+11 & 516  & 3.04E+03 & 3.65E-03     \\
0.24  & 40.3 & 9.62E+11 & 609  & 3.98E+03 & 3.15E-03     \\
0.24  & 34.5 & 1.72E+12 & 736  & 6.02E+03 & 2.95E-03     \\
\hline
0.51  & 205  & 9.78E+08 & 69.5 & 3.86E+01 & 2.05E-02     \\
0.51  & 205  & 1.78E+09 & 85.3 & 5.14E+01 & 1.40E-02     \\
0.51  & 205  & 3.14E+09 & 100  & 7.35E+01 & 9.39E-03     \\
0.51  & 206  & 5.59E+09 & 121  & 9.39E+01 & 5.80E-03     \\
0.51  & 206  & 9.91E+09 & 144  & 1.10E+02 & 4.30E-03     \\
0.51  & 175  & 1.79E+10 & 174  & 1.65E+02 & 3.21E-03     \\
\hdashline
0.24  & 205  & 9.52E+09 & 151  & 1.09E+02 & 1.08E-02     \\
0.24  & 205  & 1.73E+10 & 171  & 1.60E+02 & 6.68E-03     \\
0.24  & 205  & 3.11E+10 & 209  & 2.30E+02 & 4.93E-03     \\
0.24  & 205  & 5.57E+10 & 254  & 3.03E+02 & 3.75E-03     \\
0.24  & 204  & 9.87E+10 & 299  & 4.29E+02 & 2.62E-03     \\
0.24  & 172  & 1.77E+11 & 353  & 6.02E+02 & 1.90E-03     \\
\hline
0.24  & 993  & 3.14E+09 & 112  & 1.52E+01 & 1.62E-02     \\
0.24  & 993  & 5.59E+09 & 132  & 2.08E+01 & 1.24E-02     \\
0.24  & 994  & 9.96E+09 & 156  & 2.84E+01 & 1.02E-02     \\
0.24  & 983  & 1.74E+10 & 191  & 4.10E+01 & 6.13E-03     \\ 
\hline \hline
\end{supertabular}
\tabnote{Non-dimensional experimental values for non-rotating cases. $\varGamma=D/H$ is the aspect ratio, $Pr$ is the Prandtl number, $Ra$ is the Rayleigh number, $Nu$ is the Nusselt number, $Re_z$ is the vertical rms Reynolds number, and $\theta/\Delta T$ is the internal temperature fluctuation normalised by the vertical temperature difference. Solid horizontal lines separate the different $Pr$ fluids and dashed horizontal lines separate the different aspect ratio ($\varGamma$) cells.}
\end{center}

\begin{center}
\begin{table}[h!]
\tbl{Non-dimensional, rotating RBC data.}
{\begin{tabular}{r@{\hspace{128mm}}}
\\
\end{tabular}}
\vspace{-11mm}
\end{table}
\tablefirsthead{%
\hline
\hline
\multicolumn{1}{c}{$\varGamma$} &
\multicolumn{1}{c}{$Pr$} &
\multicolumn{1}{c}{$Ra$} &
\multicolumn{1}{c}{$E$} &
\multicolumn{1}{c}{$Nu$} &
\multicolumn{1}{c}{$Re_z$} &
\multicolumn{1}{c}{$Ro_z$} &
\multicolumn{1}{c}{$\theta/\Delta T$}\\
\hline }
\tablehead{%
\multicolumn{8}{l}{\footnotesize\sl continued from previous page}\\
\hline
\multicolumn{1}{c}{$\Gamma$} &
\multicolumn{1}{c}{$Pr$} &
\multicolumn{1}{c}{$Ra$} &
\multicolumn{1}{c}{$E$} &
\multicolumn{1}{c}{$Nu$} &
\multicolumn{1}{c}{$Re_z$} &
\multicolumn{1}{c}{$Ro_z$} &
\multicolumn{1}{c}{$\theta/\Delta T$}\\
\hline}
\tabletail{
\hline
\multicolumn{8}{r}{\footnotesize\sl continued on next page}\\
}
\tablelasttail{
\hline
}
\begin{supertabular}{cccccccc}
1.01  & 6.05 & 3.28E+08 & 3.00E-05 & 51.6 & \gr{4.67E+02} & \gr{1.40E-02} & 3.55E-02     \\
1.01  & 6.06 & 3.60E+08 & 3.00E-06 & 10.8 & \gr{4.53E+02} & \gr{1.36E-03} & 2.81E-02     \\
1.01  & 6.04 & 5.50E+08 & 2.99E-05 & 59.3 & \gr{5.07E+02} & \gr{1.52E-02} & 3.07E-02     \\
1.01  & 6.05 & 5.94E+08 & 2.99E-06 & 24.3 & \gr{4.49E+02} & \gr{1.34E-03} & 4.19E-02     \\
1.01  & 6.03 & 9.90E+08 & 2.99E-05 & 67.1 & 5.78E+02 & 1.73E-02 & 2.24E-02     \\
1.01  & 6.04 & 1.03E+09 & 2.99E-06 & 49.7 & \gr{4.87E+02} & \gr{1.46E-03} & 5.15E-02     \\
1.01  & 6.00 & 1.76E+09 & 2.97E-05 & 78.8 & 7.31E+02 & 2.17E-02 & 1.70E-02     \\
1.01  & 6.00 & 1.77E+09 & 2.98E-06 & 72.9 & 5.76E+02 & 1.71E-03 & 4.08E-02     \\
1.01  & 5.91 & 3.16E+09 & 2.93E-05 & 91.4 & 9.56E+02 & 2.80E-02 & 1.46E-02     \\
1.01  & 5.93 & 3.17E+09 & 2.94E-06 & 93.4 & 6.98E+02 & 2.06E-03 & 2.66E-02     \\
1.01  & 5.44 & 5.79E+09 & 3.00E-05 & 107  & 1.44E+03 & 4.33E-02 & 1.28E-02     \\
1.01  & 5.44 & 5.79E+09 & 3.00E-06 & 107  & 9.37E+02 & 2.81E-03 & 2.02E-02     \\
\hdashline
0.51  & 6.06 & 1.76E+09 & 3.01E-05 & 77.0 & \gr{9.75E+02} & \gr{2.94E-02} & 2.55E-02     \\
0.51  & 6.06 & 1.75E+09 & 1.00E-05 & 83.9 & \gr{9.15E+02} & \gr{9.19E-03} & 3.94E-02     \\
0.51  & 6.06 & 1.89E+09 & 1.00E-06 & 20.1 & \gr{8.07E+02} & \gr{8.11E-04} & 3.30E-02     \\
0.51  & 6.06 & 3.17E+09 & 3.01E-05 & 89.6 & 1.18E+03 & 3.54E-02 & 2.10E-02     \\
0.51  & 6.07 & 3.32E+09 & 1.01E-06 & 53.1 & \gr{8.35E+02} & \gr{8.40E-04} & 5.08E-02     \\
0.51  & 6.06 & 5.62E+09 & 3.01E-05 & 105  & 1.50E+03 & 4.53E-02 & 1.58E-02     \\
0.51  & 6.06 & 5.56E+09 & 1.00E-05 & 115  & 1.24E+03 & 1.24E-02 & 2.43E-02     \\
0.51  & 6.06 & 5.76E+09 & 1.00E-06 & 84.5 & \gr{9.16E+02} & \gr{9.20E-04} & 5.28E-02     \\
0.51  & 6.05 & 1.01E+10 & 3.01E-05 & 123  & 2.10E+03 & 6.31E-02 & 1.17E-02     \\
0.51  & 6.06 & 1.02E+10 & 1.00E-06 & 114  & \gr{1.07E+03} & \gr{1.08E-03} & 3.55E-02     \\
0.51  & 6.04 & 1.80E+10 & 3.01E-05 & 144  & 3.02E+03 & 9.06E-02 & 9.06E-03     \\
0.51  & 6.04 & 1.78E+10 & 1.00E-05 & 155  & 1.97E+03 & 1.97E-02 & 1.38E-02     \\
0.51  & 6.05 & 1.80E+10 & 1.00E-06 & 146  & 1.27E+03 & 1.27E-03 & 2.80E-02     \\
0.51  & 6.01 & 3.24E+10 & 2.99E-05 & 176  & 3.86E+03 & 1.15E-01 & 7.83E-03     \\
0.51  & 6.00 & 3.23E+10 & 9.96E-07 & 183  & 1.70E+03 & 1.70E-03 & 2.21E-02     \\
0.51  & 5.09 & 5.82E+10 & 2.58E-05 & 211  & 5.77E+03 & 1.49E-01 & 7.17E-03     \\
0.51  & 5.08 & 5.80E+10 & 8.61E-06 & 217  & 4.82E+03 & 4.15E-02 & 1.06E-02     \\
0.51  & 5.09 & 5.79E+10 & 8.61E-07 & 220  & 2.47E+03 & 2.13E-03 & 1.83E-02     \\
\hdashline
0.24  & 6.06 & 1.76E+10 & 3.00E-06 & 158  & \gr{2.17E+03} & \gr{6.51E-03} & 2.70E-02     \\
0.24  & 6.06 & 1.77E+10 & 1.00E-06 & 152  & \gr{1.93E+03} & \gr{1.93E-03} & 4.37E-02     \\
0.24  & 6.07 & 1.85E+10 & 3.00E-07 & 91.3 & \gr{1.74E+03} & \gr{5.23E-04} & 4.20E-02     \\
0.24  & 6.07 & 3.19E+10 & 3.00E-06 & 193  & 2.69E+03 & 8.07E-03 & 2.10E-02     \\
0.24  & 6.07 & 3.21E+10 & 1.00E-06 & 187  & \gr{2.37E+03} & \gr{2.37E-03} & 3.52E-02     \\
0.24  & 6.07 & 3.33E+10 & 3.00E-07 & 132  & \gr{2.01E+03} & \gr{6.03E-04} & 3.98E-02     \\
0.24  & 6.07 & 5.74E+10 & 3.00E-06 & 220  & 3.36E+03 & 1.01E-02 & 2.74E-02     \\
0.24  & 6.07 & 5.74E+10 & 1.00E-06 & 224  & 2.88E+03 & 2.88E-03 & 2.80E-02     \\
0.24  & 6.07 & 5.87E+10 & 3.00E-07 & 175  & \gr{2.22E+03} & \gr{6.67E-04} & 3.76E-02     \\
0.24  & 6.07 & 1.03E+11 & 3.01E-06 & 253  & 4.39E+03 & 1.32E-02 & 2.45E-02     \\
0.24  & 6.07 & 1.02E+11 & 1.00E-06 & 259  & 3.69E+03 & 3.69E-03 & 2.31E-02     \\
0.24  & 6.08 & 1.04E+11 & 3.01E-07 & 231  & 2.72E+03 & 8.17E-04 & 3.16E-02     \\
0.24  & 6.04 & 1.82E+11 & 2.99E-06 & 290  & 5.84E+03 & 1.75E-02 & 1.39E-02     \\
0.24  & 6.04 & 1.81E+11 & 9.97E-07 & 302  & 4.88E+03 & 4.87E-03 & 1.90E-02     \\
0.24  & 6.05 & 1.82E+11 & 2.99E-07 & 289  & 3.42E+03 & 1.02E-03 & 2.62E-02     \\
0.24  & 5.91 & 3.27E+11 & 2.93E-06 & 353  & 8.15E+03 & 2.39E-02 & 1.09E-02     \\
0.24  & 5.90 & 3.26E+11 & 9.76E-07 & 364  & 6.27E+03 & 6.13E-03 & 1.45E-02     \\
0.24  & 5.90 & 3.27E+11 & 2.93E-07 & 356  & 4.78E+03 & 1.40E-03 & 2.03E-02     \\
0.24  & 4.99 & 5.88E+11 & 2.53E-06 & 424  & 1.25E+04 & 3.17E-02 & 9.53E-03     \\
0.24  & 4.98 & 5.86E+11 & 8.41E-07 & 437  & 9.02E+03 & 7.58E-03 & 1.33E-02     \\
0.24  & 4.99 & 5.88E+11 & 2.53E-07 & 424  & 7.46E+03 & 1.89E-03 & 1.77E-02     \\
\hline
1.01  & 40.9 & 6.83E+08 & 4.41E-04 & 54.4 & \gr{1.72E+02} & \gr{7.57E-02} & 1.43E-02     \\
1.01  & 40.9 & 6.79E+08 & 1.00E-04 & 59.8 & \gr{1.40E+02} & \gr{1.40E-02} & 2.21E-02     \\
1.01  & 40.9 & 6.89E+08 & 1.00E-05 & 65.8 & \gr{1.35E+02} & \gr{1.35E-03} & 4.77E-02     \\
1.01  & 40.9 & 1.07E+09 & 4.41E-04 & 65.0 & 2.03E+02 & 8.97E-02 & 1.20E-02     \\
1.01  & 40.9 & 1.07E+09 & 1.00E-04 & 70.9 & \gr{1.59E+02} & \gr{1.59E-02} & 1.66E-02     \\
1.01  & 40.9 & 9.71E+08 & 3.00E-05 & 80.9 & \gr{1.46E+02} & \gr{4.39E-03} & 2.59E-02     \\
1.01  & 41.0 & 1.08E+09 & 1.00E-05 & 86.5 & \gr{1.40E+02} & \gr{1.40E-03} & 4.94E-02     \\
1.01  & 40.9 & 1.78E+09 & 4.41E-04 & 77.6 & 2.62E+02 & 1.15E-01 & 9.00E-03     \\
1.01  & 40.9 & 1.77E+09 & 9.99E-05 & 84.8 & \gr{1.92E+02} & \gr{1.92E-02} & 1.35E-02     \\
1.01  & 40.9 & 1.76E+09 & 3.00E-05 & 94.5 & \gr{1.71E+02} & \gr{5.14E-03} & 2.05E-02     \\
1.01  & 40.9 & 1.77E+09 & 9.98E-06 & 104  & \gr{1.52E+02} & \gr{1.52E-03} & 3.61E-02     \\
1.01  & 41.0 & 3.26E+09 & 4.42E-04 & 93.8 & 3.39E+02 & 1.50E-01 & 6.59E-03     \\
1.01  & 41.0 & 3.25E+09 & 1.00E-04 & 101  & 2.45E+02 & 2.45E-02 & 9.68E-03     \\
1.01  & 41.0 & 3.24E+09 & 3.01E-05 & 111  & 2.05E+02 & 6.15E-03 & 1.48E-02     \\
1.01  & 41.0 & 3.22E+09 & 1.00E-05 & 124  & \gr{1.73E+02} & \gr{1.73E-03} & 2.45E-02     \\
1.01  & 41.0 & 5.71E+09 & 4.42E-04 & 111  & 4.41E+02 & 1.95E-01 & 4.83E-03     \\
1.01  & 41.0 & 5.70E+09 & 1.00E-04 & 119  & 3.31E+02 & 3.32E-02 & 7.87E-03     \\
1.01  & 41.0 & 5.68E+09 & 3.01E-05 & 128  & 2.63E+02 & 7.90E-03 & 1.20E-02     \\
1.01  & 41.0 & 5.65E+09 & 1.00E-05 & 142  & 2.05E+02 & 2.05E-03 & 1.86E-02     \\
1.01  & 41.1 & 1.01E+10 & 4.43E-04 & 132  & 5.41E+02 & 2.40E-01 & 3.44E-03     \\
1.01  & 41.1 & 1.01E+10 & 1.00E-04 & 140  & 4.59E+02 & 4.60E-02 & 6.42E-03     \\
1.01  & 41.1 & 1.00E+10 & 3.02E-05 & 149  & 3.40E+02 & 1.03E-02 & 9.35E-03     \\
1.01  & 41.1 & 1.00E+10 & 1.00E-05 & 163  & 2.61E+02 & 2.62E-03 & 1.27E-02     \\
1.01  & 40.7 & 1.25E+10 & 4.39E-04 & 142  & 5.49E+02 & 2.41E-01 & 3.09E-03     \\
1.01  & 40.7 & 1.25E+10 & 2.98E-05 & 151  & 3.75E+02 & 1.12E-02 & 7.91E-03     \\
1.01  & 39.4 & 1.43E+10 & 4.25E-04 & 149  & 5.49E+02 & 2.34E-01 & 2.87E-03     \\
1.01  & 37.3 & 1.77E+10 & 4.03E-04 & 159  & 5.16E+02 & 2.08E-01 & 2.81E-03     \\
1.01  & 37.3 & 1.77E+10 & 9.14E-05 & 166  & 6.80E+02 & 6.21E-02 & 5.55E-03     \\
1.01  & 37.3 & 1.76E+10 & 9.13E-06 & 187  & 3.57E+02 & 3.26E-03 & 1.00E-02     \\
\hdashline
0.51  & 41.0 & 5.68E+09 & 1.00E-04 & 117  & \gr{3.91E+02} & \gr{3.93E-02} & 2.52E-02     \\
0.51  & 41.0 & 6.00E+09 & 3.01E-05 & 128  & \gr{3.10E+02} & \gr{9.35E-03} & 2.25E-02     \\
0.51  & 41.0 & 5.75E+09 & 1.00E-05 & 136  & \gr{2.96E+02} & \gr{2.97E-03} & 2.91E-02     \\
0.51  & 41.0 & 1.02E+10 & 1.01E-04 & 143  & 4.90E+02 & 4.92E-02 & 1.20E-02     \\
0.51  & 41.0 & 1.02E+10 & 3.01E-05 & 150  & \gr{3.74E+02} & \gr{1.13E-02} & 1.53E-02     \\
0.51  & 41.0 & 1.77E+10 & 1.00E-04 & 174  & 6.83E+02 & 6.86E-02 & 8.65E-03     \\
0.51  & 41.0 & 1.79E+10 & 3.01E-05 & 176  & 4.48E+02 & 1.35E-02 & 1.05E-02     \\
0.51  & 41.0 & 1.77E+10 & 1.00E-05 & 191  & 4.15E+02 & 4.17E-03 & 1.66E-02     \\
0.51  & 40.9 & 3.11E+10 & 1.00E-04 & 208  & 9.34E+02 & 9.38E-02 & 6.08E-03     \\
0.51  & 41.0 & 3.13E+10 & 3.01E-05 & 209  & 6.17E+02 & 1.86E-02 & 8.37E-03     \\
0.51  & 41.0 & 5.61E+10 & 1.01E-04 & 251  & 1.25E+03 & 1.25E-01 & 5.19E-03     \\
0.51  & 41.0 & 5.62E+10 & 3.01E-05 & 246  & 9.72E+02 & 2.93E-02 & 6.76E-03     \\
0.51  & 41.0 & 5.60E+10 & 1.00E-05 & 261  & 6.82E+02 & 6.85E-03 & 9.01E-03     \\
0.51  & 40.8 & 9.82E+10 & 1.00E-04 & 302  & 1.66E+03 & 1.66E-01 & 4.03E-03     \\
0.51  & 40.8 & 9.84E+10 & 3.00E-05 & 291  & 1.41E+03 & 4.24E-02 & 5.69E-03     \\
0.51  & 35.0 & 1.76E+11 & 8.62E-05 & 358  & 2.44E+03 & 2.10E-01 & 3.25E-03     \\
0.51  & 35.0 & 1.76E+11 & 2.58E-05 & 352  & 2.37E+03 & 6.11E-02 & 4.55E-03     \\
0.51  & 35.0 & 1.76E+11 & 8.62E-06 & 362  & 1.43E+03 & 1.23E-02 & 5.50E-03     \\
\hdashline
0.24  & 41.0 & 5.67E+10 & 2.50E-05 & 239  & 9.56E+02 & 2.39E-02 & 1.36E-02     \\
0.24  & 41.0 & 5.70E+10 & 1.00E-05 & 252  & 8.98E+02 & 8.98E-03 & 2.32E-02     \\
0.24  & 41.0 & 5.70E+10 & 3.00E-06 & 274  & \gr{7.54E+02} & \gr{2.26E-03} & 2.33E-02     \\
0.24  & 41.0 & 5.71E+10 & 1.00E-06 & 291  & \gr{6.88E+02} & \gr{6.88E-04} & 3.25E-02     \\
0.24  & 41.0 & 1.03E+11 & 2.50E-05 & 307  & 1.39E+03 & 3.47E-02 & 1.05E-02     \\
0.24  & 41.0 & 1.03E+11 & 1.00E-05 & 312  & 1.17E+03 & 1.18E-02 & 1.08E-02     \\
0.24  & 41.0 & 1.03E+11 & 3.00E-06 & 335  & 1.01E+03 & 3.04E-03 & 1.47E-02     \\
0.24  & 41.0 & 1.03E+11 & 1.00E-06 & 361  & \gr{8.41E+02} & \gr{8.42E-04} & 2.66E-02     \\
0.24  & 41.0 & 1.81E+11 & 2.50E-05 & 374  & 1.70E+03 & 4.25E-02 & 7.22E-03     \\
0.24  & 41.0 & 1.81E+11 & 1.00E-05 & 364  & 1.52E+03 & 1.52E-02 & 9.68E-03     \\
0.24  & 41.0 & 1.81E+11 & 3.00E-06 & 389  & 1.19E+03 & 3.59E-03 & 1.28E-02     \\
0.24  & 41.0 & 1.81E+11 & 1.00E-06 & 418  & 9.79E+02 & 9.80E-04 & 1.94E-02     \\
0.24  & 41.0 & 3.19E+11 & 2.50E-05 & 435  & 2.31E+03 & 5.77E-02 & 5.47E-03     \\
0.24  & 41.0 & 3.19E+11 & 1.00E-05 & 426  & 2.21E+03 & 2.21E-02 & 7.79E-03     \\
0.24  & 41.0 & 3.19E+11 & 3.00E-06 & 436  & 1.62E+03 & 4.85E-03 & 8.91E-03     \\
0.24  & 41.0 & 3.18E+11 & 1.00E-06 & 468  & 1.24E+03 & 1.24E-03 & 1.32E-02     \\
0.24  & 41.0 & 5.64E+11 & 2.50E-05 & 516  & 3.11E+03 & 7.78E-02 & 4.26E-03     \\
0.24  & 41.0 & 5.64E+11 & 1.00E-05 & 508  & 2.75E+03 & 2.75E-02 & 6.11E-03     \\
0.24  & 41.0 & 5.64E+11 & 3.00E-06 & 515  & 1.90E+03 & 5.69E-03 & 7.30E-03     \\
0.24  & 41.0 & 5.63E+11 & 1.00E-06 & 534  & 1.58E+03 & 1.58E-03 & 1.05E-02     \\
0.24  & 40.3 & 9.62E+11 & 2.46E-05 & 604  & 4.10E+03 & 1.01E-01 & 3.61E-03     \\
0.24  & 40.3 & 9.64E+11 & 9.84E-06 & 596  & 4.03E+03 & 3.96E-02 & 4.68E-03     \\
0.24  & 40.3 & 9.64E+11 & 2.95E-06 & 595  & 2.68E+03 & 7.92E-03 & 6.35E-03     \\
0.24  & 40.3 & 9.63E+11 & 9.84E-07 & 610  & 2.25E+03 & 2.21E-03 & 8.08E-03     \\
0.24  & 34.5 & 1.72E+12 & 2.12E-05 & 733  & 6.19E+03 & 1.31E-01 & 3.04E-03     \\
0.24  & 34.5 & 1.72E+12 & 8.46E-06 & 727  & 5.58E+03 & 4.72E-02 & 4.07E-03     \\
0.24  & 34.5 & 1.72E+12 & 2.54E-06 & 710  & 4.55E+03 & 1.15E-02 & 6.16E-03     \\
0.24  & 34.4 & 1.72E+12 & 8.46E-07 & 717  & 3.53E+03 & 2.98E-03 & 7.48E-03     \\
\hline
0.51  & 205  & 9.84E+08 & 4.42E-04 & 66.2 & \gr{4.23E+01} & \gr{1.87E-02} & 2.82E-02     \\
0.51  & 205  & 9.82E+08 & 1.00E-04 & 72.0 & \gr{3.60E+01} & \gr{3.61E-03} & 1.91E-02     \\
0.51  & 205  & 9.47E+08 & 3.02E-05 & 82.1 & \gr{3.36E+01} & \gr{1.01E-03} & 3.32E-02     \\
0.51  & 205  & 1.78E+09 & 4.43E-04 & 82.5 & \gr{5.30E+01} & \gr{2.35E-02} & 2.48E-02     \\
0.51  & 205  & 1.78E+09 & 1.01E-04 & 87.5 & \gr{4.02E+01} & \gr{4.04E-03} & 1.74E-02     \\
0.51  & 206  & 1.74E+09 & 3.02E-05 & 97.1 & \gr{3.76E+01} & \gr{1.13E-03} & 2.75E-02     \\
0.51  & 205  & 3.15E+09 & 4.43E-04 & 99.7 & 6.47E+01 & 2.86E-02 & 1.69E-02     \\
0.51  & 205  & 3.14E+09 & 1.00E-04 & 103  & \gr{4.91E+01} & \gr{4.93E-03} & 1.43E-02     \\
0.51  & 205  & 3.10E+09 & 3.02E-05 & 113  & \gr{4.49E+01} & \gr{1.35E-03} & 1.85E-02     \\
0.51  & 205  & 5.60E+09 & 4.43E-04 & 120  & 8.40E+01 & 3.72E-02 & 1.12E-02     \\
0.51  & 205  & 5.59E+09 & 1.01E-04 & 122  & 6.23E+01 & 6.26E-03 & 1.13E-02     \\
0.51  & 206  & 5.54E+09 & 3.02E-05 & 131  & \gr{5.24E+01} & \gr{1.58E-03} & 1.51E-02     \\
0.51  & 206  & 9.92E+09 & 4.43E-04 & 143  & 1.15E+02 & 5.09E-02 & 6.62E-03     \\
0.51  & 205  & 9.92E+09 & 1.01E-04 & 145  & 8.10E+01 & 8.15E-03 & 8.54E-03     \\
0.51  & 206  & 9.86E+09 & 3.02E-05 & 153  & 7.05E+01 & 2.13E-03 & 1.14E-02     \\
0.51  & 175  & 1.79E+10 & 3.78E-04 & 173  & 1.80E+02 & 6.79E-02 & 5.31E-03     \\
0.51  & 175  & 1.79E+10 & 8.59E-05 & 175  & 1.37E+02 & 1.18E-02 & 7.62E-03     \\
0.51  & 175  & 1.79E+10 & 2.58E-05 & 182  & 9.94E+01 & 2.56E-03 & 1.00E-02     \\
\hdashline
0.24  & 205  & 9.42E+09 & 9.98E-05 & 150  & \gr{1.03E+02} & \gr{1.03E-02} & 2.73E-02     \\
0.24  & 205  & 9.49E+09 & 3.00E-05 & 152  & \gr{1.01E+02} & \gr{3.02E-03} & 1.57E-02     \\
0.24  & 205  & 9.51E+09 & 9.98E-06 & 180  & \gr{8.73E+01} & \gr{8.71E-04} & 2.18E-02     \\
0.24  & 205  & 1.72E+10 & 9.98E-05 & 171  & 1.26E+02 & 1.26E-02 & 1.08E-02     \\
0.24  & 205  & 1.73E+10 & 3.00E-05 & 171  & 1.27E+02 & 3.81E-03 & 8.41E-03     \\
0.24  & 205  & 1.73E+10 & 9.98E-06 & 195  & 9.33E+01 & 9.31E-04 & 2.14E-02     \\
0.24  & 205  & 3.12E+10 & 9.99E-05 & 206  & 1.68E+02 & 1.68E-02 & 8.85E-03     \\
0.24  & 205  & 3.13E+10 & 3.00E-05 & 205  & 1.38E+02 & 4.13E-03 & 8.00E-03     \\
0.24  & 205  & 3.12E+10 & 9.99E-06 & 225  & \gr{1.20E+02} & \gr{1.20E-03} & 1.46E-02     \\
0.24  & 205  & 5.58E+10 & 9.99E-05 & 249  & 2.23E+02 & 2.22E-02 & 6.88E-03     \\
0.24  & 205  & 5.58E+10 & 3.00E-05 & 246  & 1.88E+02 & 5.63E-03 & 7.81E-03     \\
0.24  & 205  & 5.58E+10 & 9.99E-06 & 258  & 1.79E+02 & 1.79E-03 & 9.79E-03     \\
0.24  & 205  & 9.87E+10 & 9.96E-05 & 298  & 2.94E+02 & 2.93E-02 & 5.33E-03     \\
0.24  & 205  & 9.88E+10 & 2.99E-05 & 294  & 2.41E+02 & 7.22E-03 & 7.50E-03     \\
0.24  & 205  & 9.88E+10 & 9.96E-06 & 299  & 2.32E+02 & 2.31E-03 & 8.55E-03     \\
0.24  & 172  & 1.77E+11 & 8.39E-05 & 351  & 4.84E+02 & 4.06E-02 & 4.04E-03     \\
0.24  & 172  & 1.77E+11 & 2.52E-05 & 347  & 3.91E+02 & 9.85E-03 & 5.60E-03     \\
0.24  & 172  & 1.77E+11 & 8.39E-06 & 348  & 3.40E+02 & 2.86E-03 & 6.83E-03     \\
\hline
0.24  & 994  & 1.80E+09 & 4.41E-04 & 86.4 & \gr{1.45E+01} & \gr{6.41E-03} & 1.80E-02     \\
0.24  & 994  & 1.80E+09 & 1.00E-04 & 87.0 & \gr{1.43E+01} & \gr{1.43E-03} & 1.78E-02     \\
0.24  & 994  & 1.80E+09 & 3.00E-05 & 103  & \gr{1.43E+01} & \gr{4.29E-04} & 2.38E-02     \\
0.24  & 993  & 3.14E+09 & 4.40E-04 & 115  & \gr{1.78E+01} & \gr{7.85E-03} & 1.88E-02     \\
0.24  & 993  & 3.15E+09 & 1.00E-04 & 108  & \gr{1.55E+01} & \gr{1.55E-03} & 1.04E-02     \\
0.24  & 993  & 3.14E+09 & 3.00E-05 & 126  & \gr{1.56E+01} & \gr{4.68E-04} & 1.65E-02     \\
0.24  & 993  & 5.59E+09 & 4.40E-04 & 130  & \gr{1.85E+01} & \gr{8.16E-03} & 1.28E-02     \\
0.24  & 993  & 5.60E+09 & 1.00E-04 & 125  & \gr{1.74E+01} & \gr{1.74E-03} & 8.82E-03     \\
0.24  & 993  & 5.59E+09 & 3.00E-05 & 141  & \gr{1.75E+01} & \gr{5.26E-04} & 1.13E-02     \\
0.24  & 993  & 9.96E+09 & 4.41E-04 & 155  & \gr{2.20E+01} & \gr{9.70E-03} & 1.28E-02     \\
0.24  & 993  & 9.98E+09 & 1.00E-04 & 150  & \gr{2.10E+01} & \gr{2.10E-03} & 1.05E-02     \\
0.24  & 993  & 9.96E+09 & 3.00E-05 & 165  & \gr{1.86E+01} & \gr{5.60E-04} & 9.87E-03     \\
0.24  & 983  & 1.74E+10 & 4.36E-04 & 189  & 4.30E+01 & 1.87E-02 & 9.79E-03     \\
0.24  & 983  & 1.74E+10 & 9.91E-05 & 182  & 2.76E+01 & 2.74E-03 & 8.83E-03     \\
0.24  & 983  & 1.74E+10 & 2.97E-05 & 189  & \gr{2.35E+01} & \gr{6.99E-04} & 7.29E-03     \\
0.24  & 814  & 3.11E+10 & 3.65E-04 & 230  & 6.79E+01 & 2.48E-02 & 5.01E-03     \\
0.24  & 814  & 3.12E+10 & 8.28E-05 & 224  & 4.09E+01 & 3.38E-03 & 7.97E-03     \\
0.24  & 814  & 3.11E+10 & 2.48E-05 & 230  & 3.32E+01 & 8.26E-04 & 6.08E-03     \\
\hline \hline
\end{supertabular}
\tabnote{Non-dimensional experimental values for rotating cases. $\varGamma=D/H$ is the aspect ratio, $Pr$ is the Prandtl number, $Ra$ is the Rayleigh number, $E$ is the Ekman number, $Nu$ is the Nusselt number, $Re_z$ is the vertical rms Reynolds number, $Ro_z$ is the vertical rms Rossby number, and $\theta/\Delta T$ is the internal temperature fluctuation normalised by the vertical temperature difference. Solid horizontal lines separate the different $Pr$ fluids and dashed horizontal lines separate the different aspect ratio ($\varGamma$) cells. Gray coloured values of $Re_z$ and $Ro_z$ indicate velocity measurements encumbered by ambient system noise (SNR $<$ 5.5 dB).}
\end{center}


\end{document}